\documentclass[11pt]{article}
\usepackage{graphicx,amsmath,amsthm,mathrsfs,amssymb}
\usepackage[usenames]{color}
\usepackage{ulem}

\newcommand{\ee}{\end{equation}}
\newcommand{\word}[1]{\,\,\mbox{#1}\,\,}
\newcommand{\reff}[1]{(\ref{#1})}
\newcommand{\beq}{\begin{equation}}
\newcommand{\eeq}[1]{\label{#1}\end{equation}}
\newcommand{\beqa}{\begin{eqnarray}}
\newcommand{\eea}{\end{eqnarray}}
\newcommand{\eeqa}[1]{\label{#1}\end{eqnarray}}
\newcommand{\beg}{\begin{equation*}}
\newcommand{\eeg}{\end{equation*}}

\newcommand{\eh}{E\!H}
\newcommand{\kg}{K\!G}

\newcommand{\eq}{\!=\!}
\newcommand{\p}{\!+\!}
\newcommand{\m}{\!-\!}

\newcommand{\less}{\!<\!}
\newcommand{\great}{\!>\!}

\newcommand{\del}[2]{\Big(\dfrac{\partial {#1}}{\partial {#2}}\Big)}

\newcommand{\bsplit}{\begin{split}}
\newcommand{\esplit}{\end{split}}

\begin{document}
\begin{titlepage}
\title{Numerical thermodynamic studies of classical \\gravitational collapse in 3+1 and 4+1 dimensions}
\author{\thanks{bconstantine07@ubishops.ca} Benjamin Constantineau and \thanks{aedery@ubishops.ca} Ariel Edery \\\\{\small\it Physics Department, Bishop's University}\\
{\small\it 2600 College Street, Sherbrooke, Qu\'{e}bec, Canada
J1M~0C8}}
\date{}
\maketitle
\begin{abstract}
Recent numerical work on the gravitational collapse of a 5D (4+1) Yang-Mills instanton has provided numerical evidence that the free energy $F=E-TS=E/3$ of a 5D Schwarzschild black hole of mass $E$ can be obtained classically via the Lagrangian. Although there is no Hawking radiation, these numerical results suggest that the quantity $TS$ has a classical meaning. We investigate this association for the physically relevant case of 3+1 dimensional collapse. We track numerically the negative of the total Lagrangian -L during the gravitational collapse of a massless scalar field to a Schwarzschild black hole in isotropic coordinates. We show that $-L$ approaches the free energy $F=E-TS=E/2$ of a 4D Schwarzschild black hole to within 5\%. We also show that the matter contribution to the free energy tends towards zero so that the entropy at late stages of the collapse is gravitational in origin. The entropy $S$ makes a negative contribution to the free energy and this feature is observed in our numerical simulation. There is a pronounced dip (negative contribution) in a thin slice just inside the event horizon precisely where the metric field is nonstationary. This is in accord with recent work suggesting black hole entropy is connected with the nonstationary phase space hidden behind the event horizon. We also obtain thermodynamic results for the 5D collapse of a massless scalar field which confirms that previous 5D results are universal and independent of the type of matter undergoing the collapse.
\end{abstract}
\end{titlepage}
\setcounter{page}{1}
\section{Introduction}

The study of thermodynamics during gravitational collapse provides a window into the process by which black holes reach their final thermal state. The majority of the work to date has consisted of analytical and semi-analytical studies of black hole entropy and Hawking radiation during shell or dust collapse \cite{Greenwood1}-\cite{Gao}, and the approaches have been quantum mechanical or semi-classical. Recently, a classical approach, based on the free energy, was carried out \cite{Khlebnikov}. The authors conducted a numerical study of classical thermodynamics during the gravitational collapse of a 5D ($4+1$) Yang Mills instanton to a Schwarzschild black hole in isotropic coordinates\footnote{The 5D Yang Mills instanton acts like dust \cite{Volkov} so that any initial static state is guaranteed to undergo gravitational collapse to a black hole.}. The function $F\eq -dI/dt\eq -L$ was tracked during the collapse process where $I$ is the total gravitational plus matter action, $L$ is the total Lagrangian and $t$ is the time measured by a stationary clock at infinity. At late stages of the collapse, $F$ approaches a constant that can be identified with the free energy $F= E-TS$ where $E$ is the ADM mass of the Schwarzschild black hole and $S$ and $T$ are the entropy and Hawking temperature respectively (the identification with the free energy is discussed later). The entropy and temperature for a $5D$ Schwarzschild black hole are known from standard black hole thermodynamics \cite{Bekenstein, Hawking1}: $T\eq \hbar/(4\pi\,r_0)$, $S\eq 4\pi^2\,r_0^3/(\hbar\,G_5)$ where $r_0$ is the gravitational radius given by $r_0^2\eq 2\,G_5\,E/(3\,\pi)$ and $G_5$ is Newton's constant in $5D$. The product $S\,T$ is equal to $2E/3$ so that the free energy in $5D$ is equal to $E/3$. The authors showed that at late stages of the collapse the function $F\eq -L$ approached a numerical value close to $E/3$.  Although there is no Hawking radiation, the numerical results suggest that the product $TS$ can be obtained classically via the gravitational Lagrangian. Note that though $S$ and $T$ each contain $\hbar$, the product $S\,T$ appearing in the free energy does not.  

We investigate the association between $-L$ and the free energy $E-TS$ for the physically relevant case of 3+1 dimensional collapse (for completeness, we also consider a massless scalar field in 4+1 dimensions). We track numerically the function  $-L$ during the classical gravitational collapse of a $4D$ and $5D$ massless scalar field to a Schwarzschild black hole in isotropic coordinates. As far as we know, this is the first time that the negative of the Lagrangian has been tracked numerically during 3+1 dimensional collapse. The entropy and temperature of the $4D$ Schwarzschild black hole of mass $E$ is well known and given by $S \eq 4\pi G\,\,E^2/(\hbar)$ and $T \eq \hbar /(8\pi\, G E)$ \cite{Bekenstein, Hawking1}. The free energy $F\eq E\m T\,S$ is therefore equal to $E/2$.  At late stages of the collapse, when thermal equilibrium has been reached, $-L$ is observed to reach a constant numerical value that is within $5\%$ of the free energy $E/2$. We also study the gravitational collapse of a $5D$ massless scalar field. This case differs from the $5D$ Yang-Mills instanton of \cite{Khlebnikov} because a static massless scalar field may undergo gravitational collapse to a black hole or disperse depending on its initial configuration \cite{Choptuik}. In our simulation, this is governed by the value of the scale parameter $\lambda$ appearing in the initial state for the scalar field. For the case of collapse to a black hole, we obtain a value for $-L$ which is again within 5\% of the free energy ($E/3$ in 5D). This confirms that the previous 5D thermodynamic results obtained for the collapse of a Yang-Mills instanton \cite{Khlebnikov} are independent of the type of matter undergoing the collapse. Our results together with previous work \cite{Khlebnikov} provides considerable numerical evidence for associating a purely classical quantity, the negative of the gravitational Lagrangian, to the free energy $E-TS$ of a black hole.     
   
We show analytically that the exterior region of a Schwarzschild black hole by itself makes a contribution $E$ to $-L$. It follows that the interior makes a negative contribution (from -$TS$). This feature is observed in both our $4D$ and $5D$ simulations. In a thin region just behind the event horizon there is a significant dip (negative contribution) to the function $-L$. Moreover, one observes that the metric fields are nonstationary in this region (and are static everywhere else). This is in agreement with recent work \cite{Edery_Constantineau} that suggests that black hole entropy is connected with the nonstationary phase space hidden behind the event horizon. It is also in agreement with the entropy obtained via explicit counting of internal microstates within the context of canonical quantization of the Schwarzschild \cite{Vaz1} and the Reissner-Nordstr\"{o}m black hole \cite{Vaz2}. That the entropy of a black hole is connected to its interior is in line with Bekenstein's original view that black hole entropy is a measure of an outside observer's ignorance of the internal configurations hidden behind the event horizon \cite{Bekenstein}.  

\subsection{Metric in isotropic coordinates}
We work in isotropic coordinates, where a spherically symmetric time-dependent $4D$ metric takes the form  
\beq
ds^2= -N(r,t)^2 dt^2 +\psi(r,t)^4 (dr^2 + r^2 d\Omega^2)\,. 
\eeq{isometric}
$N$ is called the lapse function and we refer to $\psi$ as the conformal factor. The coordinate time $t$ is the time measured by a stationary clock at infinity. During our simulation of gravitational collapse, $N$ and $\psi$ remain finite and continuous everywhere and the metric \reff{isometric} covers the entire spacetime (inside and outside the event horizon). We assume asymptotic flatness so that $N$ and $\psi$ approach unity at infinity.  Black hole formation will be interpreted to coincide with the formation of an apparent horizon. This occurs when $N$ crosses zero; light cannot reach a point where $N\eq0$ from any other point in a finite time \cite{Khlebnikov, Finelli}.   
   
The metric in the {\it exterior} region of the Schwarzschild black hole in isotropic coordinates is \cite{Buchdahl, Wald3}
\beq
ds^2=-\dfrac{(1-GM/2r)^2}{(1+ GM/2r)^2}\,dt^2 +\Big(1+\dfrac{GM}{2r}\Big)^4(dr^2+r^2\,d\Omega^2)\,.
\eeq{isotropic}
The event horizon is located at $r\eq GM/2$. The region $r\great GM/2$ of metric \reff{isotropic} covers the exterior region of the Schwarzschild black hole. However, the region $r \less GM/2$ covers the exterior region a second time, not the interior, so that metric \reff{isotropic} represents a double covering of the static exterior region. Metric \reff{isotropic} is obtained via the coordinate transformation $r'\eq r\,(1+ GM/2r)^2$ where $r'$ is the radius that appears in the standard Schwarzschild form $ds^2\eq  -A(r')\,dt^2 + B(r')\,dr'^2 + r'^2 \,d\Omega^2$ with $A(r')\eq  1-2GM/r'$ and $B(r')\eq A(r')^{-1}$. The minimum value of $r'$ is $2GM$ so that the interior region $r'\less 2GM$ is not covered. This is to be expected because the metric \reff{isotropic} has a timelike hypersurface-orthogonal Killing vector in the region $r\less GM/2$ and is therefore static in that region. In contrast, all the Killing vectors in the interior region of the Schwarzschild black hole are spacelike and the region is nonstationary. To cover the interior region of the Schwarzschild black hole in isotropic coordinates, the metric coefficients in the region $r\less GM/2$ need to be time-dependent \cite{Buchdahl}.  

At late stages of the collapse process, the metric \reff{isometric} approaches the static form \reff{isotropic} only in the exterior region $r\great GM/2$. In the interior region ($r\less GM/2$), the metric functions $N(r,t)$ and $\psi(r,t)$ are time-dependent at late times in accord with the nonstationary nature of the Schwarzschild interior. The areal radius is $R\eq \psi^2\,r$ and the Schwarzschild central singularity at $R\eq 0$ corresponds in isotropic coordinates to $\psi(r,t)\to 0$ in the interior region as $t\to \infty$. The function $\psi(r,t)$ is continuous everywhere so that the interior and exterior values must match at the event horizon $r\eq GM/2$ (where the analytical solution yields $\psi \eq 2$). Therefore, as one crosses the event horizon from the exterior towards the interior, $\psi$ must decrease from a value of $2$ at the event horizon to a value approaching zero inside. This is observed in our numerical simulation.  At late times, $\psi$ has a small constant value throughout the interior except for a small region just inside the event horizon where it increases sharply to a value close to $2$. In this small region, $\psi$ is nonstationary whereas it is basically static everywhere else. We will see that the entropy of the black hole originates from the nonstationary region just inside the event horizon.     

\subsection{The free energy} 
The thermodynamic function tracked during our simulation is $F(t)\eq -dI/dt \eq-L$. It was argued in \cite{Khlebnikov} that
if $F(t)$ approaches a constant at late times in the collapse process, then it can
be identified with the free energy $F \eq E -TS$ of the black hole. The argument
is as follows: if $F(t)$ approaches a constant at late times, then the action $I$
becomes linear in time and $I \eq -F\,(t -t_0)$ in the time interval between $t_0$
and $t$. The corresponding Euclidean action $I_E$ is obtained from the
Lorentzian action $I$ via the relation $I_E \eq i I_{t\to i\tau}\eq  F\,(\tau-\tau_0)$ where $\tau$ is the Euclidean time.
The quantum partition function $Z$ is equal to the Euclidean path integral \cite{Hawking2} with time interval $\tau-\tau_0 \eq 1/T$ where $T$ is the temperature and with
fields $\phi$ obeying the periodic boundary condition $\phi(\tau_0) \eq \phi(\tau_0 +1/T)$. For
a single (classical) trajectory, the
partition function reduces to $Z \eq e^{-I_E} \eq e^{-F/T}$. We see
that $F$ is indeed the free energy $F \eq -T \ln Z$. The negative of
the Lagrangian is therefore identified with the free energy $F \eq E - TS$ at late
times. This identification was used in the numerical work of \cite{Khlebnikov} where $-L$ at late times 
was compared to $F \eq E/3$ for the case of spherical collapse of a Yang Mills instanton to a 
$5D$ Schwarzschild black hole of mass $E$. 
\section{$4D$ thermodynamics during classical gravitational collapse}
\subsection{Equations of motion in isotropic coordinates}
In this section we obtain the equations of motion for the metric and matter fields in isotropic coordinates. We consider a massless scalar field $\chi$ coupled to gravity in $4D$. The Klein-Gordon action for a massless scalar field is given by
	\beq
	I_{\kg} = -\frac{1}{2} \,\int \sqrt{-g}\,g^{\mu\nu}\partial_{\mu}\chi\,\partial_{\nu}\chi\ d^4x = \int \sqrt{-g}\,\mathcal{L}_{\kg}\, d^4x 
	\eeq{EHKG}
 where $\mathcal{L}_{\kg}\equiv (-1/2)g^{\mu\nu}\,\partial_{\mu}\chi\,\partial_{\nu}\chi$. 	The action for gravity is $I_G=I_{\eh}+I_B$ where $I_{\eh}=\frac{1}{16\pi G} \int \sqrt{-g}\,R\ d^4x$ is the Einstein-Hilbert action and $I_B$ is a boundary term that is required if the variation of the total action $I= I_{G}+I_{\kg}$ with respect to the metric is to reproduce the Einstein field equations \cite{Poisson}. We are assuming spherical symmetry so that	$\chi = \chi(r,t)$. 
 
Variation of the action $I$ with respect to the metric leads to the Einstein equations, $G_{\mu\nu}=8\pi\,G\, T_{\mu\nu}$ where $G_{\mu\nu}$ is the Einstein tensor and $T_{\mu\nu}$ is the energy-momentum tensor given by  
	\beq
		T_{\mu\nu} \eq -2 \,\del{\mathcal{L}_{\kg}}{g^{\mu\nu}} + \mathcal{L}_{\kg}\,g_{\mu\nu}\,.
	\eeq{4DScalarStressEnergyTensor}
The non-zero components of $T_{\mu\nu}$ and $G_{\mu\nu}$ for the isotropic metric \reff{isometric} are
	\vspace{-.5em}
	\begin{flushleft}{\small}
	$T_{tt} = \frac{N^2}{2} \Big( \frac{\chi'^2}{\psi^4} + \frac{\dot{\chi}^2}{N^2} \Big)\quad;\quad	
		T_{rr} = \frac{\psi^4}{2} \Big( \frac{\chi'^2}{\psi^4} + \frac{\dot{\chi}^2}{N^2} \Big)\quad;\quad 
		T_{rt} = T_{tr} = \dot{\chi}\chi'$
	
	\vspace{-0.1em}
	$T_{\theta\theta} = -\frac{\psi^4r^2}{2} \Big( \frac{\chi'^2}{\psi^4} - \frac{\dot{\chi}^2}{N^2} \Big) \quad;\quad
		T_{\phi\phi} = \sin^2\theta \ T_{\theta\theta}$
	
	\vspace{-0.1em}
	$G_{tt} = \frac{4}{r\psi^5} ( 3r\dot{\psi}^2\psi^3 - 2\psi'N^2 - r\psi''N^2 )$
	
	\vspace{-0.1em}
	$G_{rr} = \frac{2}{r\psi^2N^3} ( 2r\psi'^2N^3 + 2\psi'N^3\psi - 4r\dot{\psi}^2N\psi^4 + 2r\dot{N}\dot{\psi}\psi^5 - 2r\ddot{\psi}N\psi^5 + 2rN'\psi'N^2\psi + N'N^2\psi^2 )$
	
	\vspace{-0.1em}
	$G_{rt} = G_{tr} = \frac{4}{\psi^2N} ( \dot{\psi}\psi'N - \dot{\psi}'N\psi + \dot{\psi}N'\psi )\quad;\quad 
		G_{\phi\phi} = \sin^2\theta \,G_{\theta\theta}$
	
	\vspace{-0.1em}	
	$G_{\theta\theta} = \frac{-r}{\psi^2N^3}\,( -2\psi'N^3\psi - 2r\psi''N^3\psi + 8r\,\dot{\psi}^2N\psi^4$
	\vspace{-0.1em}
	$\qquad \qquad \qquad+ 2r\psi'^2N^3 - 4r\dot{N}\dot{\psi}\psi^5 + 4rN\psi^5\ddot{\psi}  - N'N^2\psi^2 - rN''N^2\psi^2 ) \,.$
	\end{flushleft}
	\vspace{-1em}	
		\subsubsection{Energy and momentum constraint and evolution equation for $\psi$}
		The Einstein equation $G_{tt} = \kappa^2\,T_{tt}$ where $\kappa^2 \equiv 8\pi G$ yields  
		\beq
			\frac{4}{r\psi^5} ( 3r\dot{\psi}^2\psi^3 - 2\psi'N^2 - r\psi''N^2 ) = \frac{\kappa^2N^2}{2} \Big( \frac{\chi'^2}{\psi^4} + \frac{\dot{\chi}^2}{N^2} \Big).
		\eeq{}
		The above equation is the energy constraint and it can be expressed in the convenient form   
		\beq
			- \frac{4}{\psi^5}\,\nabla^2\psi = \kappa^2\mathcal{E} - \frac{K^2}{3}\,.
		\eeq{4DScalarEConstraint}
		In \reff{4DScalarEConstraint}, $\nabla^2\psi$ is the flat space Laplacian given by
		\beq
			\nabla^2\psi = \frac{1}{r^2}\partial_r ( \psi'r^2 ) = \frac{2\psi'}{r} + \psi'' ,
		\eeq{4DScalarLaplacian}
		and $K$ is the trace of the extrinsic curvature given by \cite{Poisson}
		\beq
			K = h^{ab}K_{ab} = \frac{-1}{2N}\,h^{ab}\partial_th_{ab} = -\frac{6 \dot{\psi}}{N\psi}
		\eeq{4DScalarTrace}
		where $h_{ab}$ is the three-metric induced on a spacelike hypersurface $\Sigma_t$ at an instant of time $t$. The term $\mathcal{E}$ is interpreted as the energy density of the matter field $\chi$:
		\beq
			\mathcal{E}= \frac{1}{2} \Big( \frac{\chi'^2}{\psi^4} + \frac{\dot{\chi}^2}{N^2} \Big).
		\eeq{4DScalarEDensityA}
	The evolution equation for the field $\psi$ is given by \reff{4DScalarTrace} 
		\beq
			\frac{\dot{\psi}}{N} = -\frac{K\psi}{6}\,.
		\eeq{4DScalarEqPsi}
If $K$, $N$ and $\psi$ are known at time $t$, $\psi$ can be evaluated at the next time step. 	The energy constraint equation \reff{4DScalarEConstraint} is not an evolution equation. It is used to monitor the accuracy of the simulation.   
		
		The Einstein equation $G_{rt} = \kappa^2\,T_{rt}$ yields the momentum constraint equation
		\beq
			\frac{4}{\psi^2N} \,( \dot{\psi}\psi'N - \dot{\psi}'N\psi + \dot{\psi}\,N'\psi ) = \kappa^2 \dot{\chi}\,\chi'.
		\ee
			Using \reff{4DScalarEqPsi}, this can be expressed in the more compact form 
		\beq
			\frac{K'}{3} = \frac{\kappa^2}{2}\,\frac{\dot\chi}{N}\,\chi'.
		\eeq{4DScalarMConstraintA}
	
	\subsubsection{The evolution equation for K}
		The Einstein equation $G_{rr} = \kappa^2\,T_{rr}$ yields
		\beqa
		& \!\!\!\!\!\!& \frac{2}{r\psi^2N^3} \,(\, 2\,r\,\psi'^2 \,N^3 - 4r\dot{\psi}^2N\psi^4 
				+ 2\psi'N^3\psi + 2r\dot{N}\dot{\psi}\psi^5 - 2r\ddot{\psi}N\psi^5 
				+ 2rN'\psi'N^2\psi + N'N^2\psi^2 \,) \nonumber \\
		& \!\!\!\!\!\!&	\eq \,\frac{\kappa^2 \psi^4}{2} \Big( \frac{\chi'^2}{\psi^4} + \frac{\dot{\chi}^2}{N^2} \Big)\,.
		\eeqa{Grr}
Using the expression for the energy density $\mathcal{E}$ given by \reff{4DScalarEDensityA} and the expression for $K$ given by \reff{4DScalarEqPsi}, the above equation reduces to a time evolution equation for K,
		\beq
			\frac{\dot{K}}{N} = \frac{K^2}{2} + \frac{3}{2}\kappa^2\mathcal{E} - 6\frac{\psi'}{\psi^5} \Big( \frac{\psi'}{\psi} + \frac{1}{r} \Big) - 3\frac{N'}{N\psi^4} \Big( \frac{2\psi'}{\psi} + \frac{1}{r} \Big).
		\eeq{4DScalarEqKA}
Following a prescription by Finelli and Khlebnikov \cite{Finelli}, one can improve the numerical stability of equation \reff{4DScalarEqKA} by adding to it the term $-\frac{3}{2}w (\kappa^2\mathcal{E} - \frac{K^2}{3} + 4\frac{\nabla^2\psi}{\psi^5})$) where $w$ is a constant. The added term is equal to zero because of \reff{4DScalarEConstraint} and therefore does not alter the equation. The final evolution equation for $K$ is
	\beq
		\frac{\dot{K}}{N} = \frac{K^2}{2} (1+w) + \frac{3}{2}\kappa^2\mathcal{E} (1-w) - 6\frac{\psi'}{\psi^5} \Big( \frac{\psi'}{\psi} + \frac{1}{r} \Big) - 3\frac{N'}{N\psi^4} \Big( \frac{2\psi'}{\psi} + \frac{1}{r} \Big) - 6w\frac{\nabla^2\psi}{\psi^5}\,.
	\eeq{4DScalarEqK}
In our simulation, $w$ is set to unity.
	
	\subsubsection{Lapse function N: ordinary differential equation}
There are two metric functions that appear in the isotropic metric \reff{isometric}: $\psi(r,t)$ and $N(r,t)$. The function $\psi(r,t)$ is a dynamical variable and its evolution is determined by \reff{4DScalarEqPsi} with $K$ evolving via \reff{4DScalarEqK}. The lapse function $N(r,t)$ is not a dynamical variable and there is no explicit evolution equation for it. Nonetheless, each time step yields a new spacelike hypersurface $\Sigma_t$ and one must readjust the value of $N(r,t)$ at each time step. This is governed by the Einstein equation $G_{\theta\theta} =\kappa^2 \,T_{\theta\theta}$ i.e. 
		\beqa
			& &\frac{1}{\psi^2\,N^3\,r} \,(\, -2\psi'N^3\psi - 2r\psi''N^3\psi 
			 + 8r\dot{\psi}^2N\psi^4 + 2r\psi'^2N^3 \nonumber\\& &\quad\quad- 4r\dot{N}\dot{\psi}\psi^5 + 4rN\psi^5\ddot{\psi}
			 - N'N^2\psi^2 - rN''N^2\psi^2 \,)\eq \frac{\kappa^2\,\psi^4}{2} \Big( \frac{\chi'^2}{\psi^4} - \frac{\dot{\chi}^2}{N^2} \Big). 
		\eea
The sum of the terms with time-derivatives can be replaced by spatial derivatives via equation \reff{Grr}. The lapse function $N$ is then governed by an ordinary differential equation containing only derivatives with respect to the variable $r$,   		
		\beq
			\frac{2r}{\psi^2}\partial_r \Big( \frac{\psi'}{r\psi^3} \Big) + \frac{r}{N}\partial_r \Big( \frac{N'}{r\psi^4} \Big) = -\kappa^2\frac{\chi'^2}{\psi^4}.
		\eeq{4DScalarEqN}
		
	\subsubsection{Evolution equations for the matter field}
		Variation of the action $I_{\kg}$ given by \reff{EHKG} with respect to the massless scalar field $\chi$ yields the Klein-Gordon equation in curved spacetime, 
		\beq
			g^{\mu\nu}\,\nabla_\mu\nabla_\nu \,\chi= 0 \Rightarrow\partial_{\mu}(\sqrt{-g}\,g^{\mu\nu}\,\partial_{\nu}\chi)=0\,.
		\eeq{KG2}
		For the metric \reff{isometric}, $\sqrt{-g}= N\,\psi^6\,r^2\,\sin(\theta)$ and the above equation becomes
		\beq\dfrac{1}{r^2}\partial_{r}(N\,\psi^2\,r^2\,\chi') = \partial_{t}(\psi^6\,\dot{\chi}/N)\,.
		\eeq{chifield}
		Defining a new function, $p(r,t)$ as
		\beq
			p \equiv \psi^6\frac{\dot{\chi}}{N}
		\eeq{4DScalarEqChi}
equation \reff{chifield} can be expressed as
		\beq
			\frac{\dot{p}}{N} = \psi^2 \Big( \chi'' + \frac{N'}{N}\chi' + \frac{2\psi'}{\psi}\chi' + \frac{2}{r}\chi' \Big).
		\eeq{4DScalarEqP}
		
The evolution equation for $\chi$ is given by \reff{4DScalarEqChi} with the function $p(r,t)$ evolving according to \reff{4DScalarEqP}. We now express the energy density \reff{4DScalarEDensityA} and the momentum constraint equation \reff{4DScalarMConstraintA} in terms of $p$, i.e.
		\beq
			\mathcal{E} = \frac{1}{2} \Big( \frac{\chi'^2}{\psi^4} + \frac{p^2}{\psi^{12}} \Big),
		\eeq{4DScalarEDensity}
		and
		\beq
			\frac{K'}{3} = \frac{\kappa^2}{2\,\psi^6}\,\chi'\,p.
		\eeq{4DScalarMConstraint}
		From the form of \reff{4DScalarEDensity}, we see that $p$ acts like a momentum conjugate to the field $\chi$.  
		
To summarize, the evolution equations for $\psi$ and $K$ governing the metric field are given by \reff{4DScalarEqPsi} and \reff{4DScalarEqK}  respectively whereas the evolution equations for $\chi$ and $p$ governing the matter field are given by \reff{4DScalarEqChi} and \reff{4DScalarEqP} respectively. The lapse function $N$ does not obey an evolution equation but is evaluated at every time step via the ordinary differential equation \reff{4DScalarEqN}. Once the boundary conditions and initial states are specified, the time evolution of all the fields is unique.

\subsection{Expression for the ADM mass}
During the gravitational collapse process the total energy (the ADM mass) remains constant and this allows us to monitor the accuracy of the simulation. We therefore need to derive an expression for the ADM mass in isotropic coordinates \reff{isometric}. The ADM mass of an asymptotically flat spacetime is defined as \cite{Poisson}
\beq
	M_{ADM} = -\frac{1}{8\pi} \lim_{S_t \rightarrow \infty} \oint_{S_t} (k-k_0) \sqrt{\sigma} \, d^2\theta,
\eeq{ADMmass}
where $S_t$ is the two-sphere at spatial infinity, $\sigma_{AB}$ is the metric on $S_t$, $k = \sigma^{AB}k_{AB}$ is the trace of the extrinsic curvature of $S_t$ embedded in $\Sigma_t$, the three-dimensional spacelike hypersurface at constant $t$, and $k_0$ is the trace of the extrinsic curvature of $S_t$ embedded in flat spacetime. The trace $k$ is given by  $k\eq \nabla_a \,r^a$ \cite{Poisson} where $r^a$ is the unit vector normal to the two-sphere boundary $S_t$. In $4D$ isotropic coordinates, a quick calculation yields  $k \eq 4\psi' / \psi^3 + 2/(\psi^2\,r)$, $k_0 = 2/(\psi^2\,r)$ so that $k-k_0 \eq 4\psi'/\psi^3$. With $\sqrt{\sigma}\, d^2\theta = \psi^4\,r^2\sin\theta\,d\theta\,d\phi$, the ADM mass \reff{ADMmass} reduces to   
	\beq
		M_{ADM} = -\frac{2 r^2}{G} \psi' \psi \,\big|_{r=R} = -\frac{2 R^2}{G} \psi' \big|_{r=R} 
		\eeq{4DScalarADMMass}
where $R$ is assumed large (infinite limit) and we used the fact that $\psi \rightarrow 1$ as $R \rightarrow \infty$. Using the energy constraint equation \reff{4DScalarEConstraint}, one can express the ADM mass \reff{4DScalarADMMass} in the integral form 
	\beq
		M_{ADM} = E_{tot}=\frac{\kappa^2}{2G} \int^R_0 \Big(\mathcal{E} - \frac{K^2}{3\kappa^2} \Big)\psi^5\,r^2\, dr
	\eeq{4DScalarEtotInt}
where $K$ and $\mathcal{E}$ are given by \reff{4DScalarTrace} and \reff{4DScalarEDensityA} respectively. The ADM mass represents the total energy $E_{tot}$ of matter plus gravitation. During the collapse process, it remains a constant. The black hole mass $E$ is obtained via the relation $r_0= G\,E/2$, where $r_0$ is the location where the lapse function $N$ crosses zero at late times.	The black hole mass $E$ will be less than the ADM mass $E_{tot}$ because in the numerical simulation part of the total energy appears in an outgoing matter wave. 

\subsection{Initial states and boundary conditions}
We choose a static initial state, where $K$ and $p$ are zero. From \reff{4DScalarEqPsi} and \reff{4DScalarEqChi}, this implies that $\dot{\psi} = \dot{\chi} = 0$ at time $t=0$. For this initial state, the momentum constraint \reff{4DScalarMConstraint} is automatically satisfied. The initial field configuration for the scalar field $\chi$ is chosen to be
	\beq
		\chi(t=0,r) = \frac{8\lambda^2r^4}{(\lambda^2 + r^2)^4},
	\eeq{4DScalarIEChi}
where $\lambda$ is a scale parameter. $\chi$ decreases rapidly to zero at the two extreme limits (i.e. as $r\to 0$ and $r\to \infty$), reaches a maximum at $r \eq \lambda$ and has a half-width of $\approx 1.3\,\lambda$. The initial energy density is concentrated in a shell of approximately this width and its peak value increases as $\lambda$ decreases. For $\lambda$ sufficiently small, the self-gravitational pull becomes large enough to initiate a gravitational collapse to a black hole.
	
The initial state for the conformal factor $\psi$ is obtained by solving the energy constraint \reff{4DScalarEConstraint}. Asymptotically, the spacetime is flat: $\psi \to 1$ and $N\to 1$. To solve for $\psi$ with this boundary condition, we express it as a power expansion in the parameter $\zeta \eq \kappa^2$ i.e. 
\beq
\psi = 1 + \zeta\psi_1 + \zeta^2\psi_2 + \ldots\,.  
\eeq{ExpPsi}
The actual expansion parameter turns out to be $\zeta$ divided by some power of the scale parameter $\lambda$. By substituting the energy density $\mathcal{E}$ given by \reff{4DScalarEDensity} and the $4D$ flat-space Laplacian \reff{4DScalarLaplacian}, along with the static assumption ($K\eq p\eq 0$) into the energy constraint equation \reff{4DScalarEConstraint}, one obtains
	\beq
		\frac{1}{r^2} \partial_r ( r^2\psi') =  -\frac{\kappa^2}{8}\chi'^2\psi.
	\eeq{4DScalarIEConstraint}
	
Substituting the expansion \reff{ExpPsi} into \reff{4DScalarIEConstraint} and matching the left hand side with the right hand side  order by order in $\zeta$ we obtain
	\beqa
		\psi & \eq & 1 + \dfrac{\zeta}{\lambda^3}\frac{1}{40320\,r\,(r^2 + \lambda^2)^8} \Big[ 1575\,r^{15}\lambda + 12075\,r^{13}\lambda^3 + \ldots + 1575\,(r^2 + \lambda^2)^8 \tan^{-1}(r/\lambda) \Big] \nonumber\\  
		 && {}+ \dfrac{\zeta^2}{\lambda^8}\frac{1}{10538886758400} \Big[ 2010133125\pi^2 - \frac{2}{r(r^2+\lambda^2)^{16}} \Big( -r\lambda^2 \big( 2967339375\,r^{30} \nonumber\\
		& & {} + 47828405625\,r^{28}\lambda^2 + \ldots + 60640845881\,r^2\lambda^{28} + 4113561991\,\lambda^{30} \big) \nonumber\\
		& & {} + 1276275\,\lambda ( r^2 + \lambda^2)^8 \big( 825\,r^{16} + 4500\,r^{14}\,\lambda^2 + \ldots + 15436\,r^2\lambda^{14} + 2717\lambda^{16} \big) \tan^{-1}(r/\lambda) \nonumber\\
		& & {} + 4020266250\,r\,(r^2 + \lambda^2)^{16} [\tan^{-1}(r/\lambda)]^2 \Big) \Big] \nonumber\\
		& & {} + \ldots \nonumber
	\eeqa{4DScalarExpansionPsi}
While we only show here the first two order terms in the expansion, our results are obtained to six orders. The initial state for $N$ is obtained numerically by solving its associated ODE \reff{4DScalarEqN} for the initial values of $\psi$ and $\chi$ defined above. Starting at the outer boundary $r\eq R$, where the spacetime is flat and $N\eq1$, we iterate backwards to obtain $N$. 

Boundary conditions need to be imposed at $r\eq0$ to ensure regularity of the solution. To ensure asymptotic flatness, we also impose boundary conditions at ``infinity", the computational outer boundary $r\eq R$. The regularity conditions are $\chi'(0,t)\eq0$ and $K'(0,t)\eq0$. The boundary conditions imposed at the outer boundary are $N(R,t)\eq 1$, $K'(R,t)\eq 0$ and $p'(R,t)\eq 0$. Together with the initial matter and metric states, these lead to a unique evolution.   

\subsection{Expression for the total Lagrangian in isotropic coordinates}

We track the function $F\eq-L$ where the total Lagrangian $L$ is a sum of the Klein-Gordon and gravitational Lagrangian. The Klein-Gordon (KG) matter Lagrangian is given by
	\beq
	\begin{split}
		L_{\kg} &= \int \sqrt{-g}\,\mathcal{L}_{\kg}\, d^3x= -\dfrac{1}{2}\int \sqrt{-g}\, g^{\mu\nu}\,\partial_{\mu}\chi\,\partial_{\nu}\chi \,d^3x\,\\
&= -2\,\pi	\int_0^R N\,\psi^6\,r^2 \Big( \frac{\chi'^2}{\psi^4} - \frac{\dot{\chi}^2}{N^2} \Big)\, dr\,.
	\end{split}
	\eeq{4DScalarKGLagrangian}
where $R$ is the radius at the outer boundary. The gravitational action $I_G$ is a sum of the Einstein-Hilbert action plus a boundary term $I_B$ and a nondynamical term $I_0$ \cite{Poisson}:
	\beq
		I_{G} = \frac{1}{16\pi G} \int \sqrt{-g}\,R \, d^3x\, dt + I_B +I_0 = \int L_G \,dt 
	\ee
where $L_G$ is the gravitational Lagrangian and contains only first derivatives of the metric functions. The Ricci scalar in isotropic coordinates is given by    
	\beqa
		R &=& \frac{2}{rN^3\psi^5} \big( 18r\dot{\psi}^2\,N\psi^3 + 6r\ddot{\psi}N\psi^4 - 8\psi'N^3 - 4r\psi''N^3 \\
		& & {} - 6r\dot{N}\dot{\psi}\psi^4 - rN''N^2\psi - 2rN'\psi'N^2 - 2N'N^2\psi \big)\nonumber\,.
	\eea
$L_G$ is obtained by integrating out by parts the second derivative terms appearing in $R$ \cite{Khlebnikov}. This yields     
	\beq
		L_{G} = \frac{1}{4 G} \int_0^R \Big( 8r^2\psi'N'\psi + 8r^2N\psi'^2 - \frac{24r^2\dot{\psi}^2\psi^4}{N} \Big)\, dr.
	\eeq{4DScalarEHLagrangian}
The total Lagrangian, $L\eq L_G+L_{\kg}$, is finally given by the expression
	\beq
		L = \frac{1}{4G} \int_0^R \Big( 8r^2\,\psi'\,N'\,\psi + 8\,r^2\,N\,\psi'^2 - \frac{24\,r^2\dot{\psi}^2\psi^4}{N} \Big)\, dr - 2\,\pi\int_0^R N\,\psi^6\,r^2 \Big( \frac{\chi'^2}{\psi^4} - \frac{\dot{\chi}^2}{N^2} \Big)\, dr\,.
	\eeq{4DScalarLtot}
 
\subsubsection{Static exterior and nonstationary interior contributions to the free energy} 

We now show that for a static scenario where $\dot{\psi}\eq\dot{\chi}\eq0$ everywhere, $F\eq -L$ reduces to the mass $E$ so that the product $T\,S$ appearing in $F\eq E \m T\,S$ is zero. Setting $\dot{\psi}$ and $\dot{\chi}$ to zero in the total Lagrangian \reff{4DScalarLtot} yields
	\beq
	\begin{split}
-L_{static} & = -\dfrac{1}{4G} \int_0^R \Big( 8r^2\,\psi'\,N'\,\psi + 8\,r^2\,N\,\psi'^{\, 2} \Big)dr + 2\,\pi\int_0^R N\,\psi^2\,r^2\,\chi'^{\, 2}\, dr\\
&= -\frac{2r^2}{G}\psi'\,N\,\psi\lvert^R_{_0} +\dfrac{1}{4G} \int_0^R\Big( 16 \,r\psi'N\psi+ 8\,r^2\,\psi''\,N\,\psi \Big)dr + 2\,\pi\int_0^R N\,\psi^2\,r^2\,\chi'^{\, 2}\, dr\\
&=-\frac{2R^2}{G}\psi'\lvert_{r=R}\,\,\,\,= \,E
\end{split}
\eeq{LStatic}
where we integrated by parts and used the energy constraint \reff{4DScalarEConstraint} with $\dot{\psi}=\dot{\chi}=0$. The last line is simply the expression for the ADM mass (note that asymptotic flatness implies that $N\eq\psi\eq 1$ at $r\eq R$).    

The Schwarzschild black hole is not static everywhere: it is not static in the interior region, inside the event horizon. Its free energy will therefore not reduce to $E$ as above. This agrees with the fact that the Schwarzschild black hole has a nonzero temperature $T$ and a nonzero entropy $S$ so that $T\,S \ne 0$. Nonetheless, we can still show that the static exterior region ($r \great GM/2$) contributes by itself $E$ to the free energy. It then follows that the $-T\,S$ contribution stems from the nonstationary interior. The metric \reff{isotropic}, which describes the exterior region of the Schwarzschild black hole, corresponds to a conformal factor of $\psi(r)\eq 1+ GM/2r$ and lapse function $N(r)\eq (1-GM/2r)/(1+ GM/2r)$. The Schwarzschild exterior is a vacuum where $\chi'$ and $\dot{\chi}$ are zero. Substituting these values into the Lagrangian \reff{4DScalarLtot} and integrating from the event horizon at $r\eq GM/2$ to infinity yields
\beq
-L_{exterior} =  -\dfrac{1}{4G} \int_{GM/2}^{\infty} \Big( 8r^2\,\psi'\,N'\,\psi + 8\,r^2\,N\,\psi'^{\, 2} \Big)\,dr = M
\eeq{Lexterior}
where $M$ and $E$ are the same quantities (we work in units where $c\eq1$). The free energy $F\eq E-T\,S$ of a Schwarzschild black hole separates nicely into two contributions: a positive contribution $E$ from the static exterior region and a negative contribution $-T\,S$ from the nonstationary interior\footnote{It should in principle be possible to calculate analytically the contribution of the interior. However, this would be a lengthy calculation more appropriate for a separate analytical study of the free energy of stationary black holes.}. This feature of a Schwarzschild black hole is observed in our numerical simulation. At late stages of the collapse, the metric field $\psi$ is nonstationary only in a thin slice just behind the event horizon. This small nonstationary region makes a negative contribution to the free energy (a large dip is observed just behind the event horizon in the accumulation plot of the free energy). The exterior is basically static and makes a positive contribution.

That black hole entropy stems from the nonstationary interior region is discussed in ref. \cite{Edery_Constantineau}. Black hole entropy is a measure of an outside observer's ignorance of the internal configurations hidden behind the event horizon \cite{Bekenstein}. In \cite{Edery_Constantineau}, the internal configurations were identified as points in phase space, the classical microstates $[h_{ab},P^{ab}]$ where $h_{ab}$ is the three-metric in the usual $3+1$ decomposition \cite{Poisson} and $P^{ab}$ is its momentum conjugate. A nonstationary interior implies that $h_{ab}$ is time-dependent and that there is a continuous set of classical microstates $[h_{ab}(t),P^{ab}(t)]$ parametrized by the time $t$. The outside observer cannot track the parameter $t$ and is clearly ignorant of which classical microstate the black hole interior is in. This implies a non-zero black hole entropy. 
	
\subsection{4D numerical results}
We work in geometrized units where $G \eq c\eq 1$. In these units, $\kappa^2\eq
8\pi$ and mass, energy, time and distance have dimensions of length. In our
simulation, these quantities are expressed in units of the ADM mass $E_{tot}$
which is normalized to unity. The scale parameter used in our simulation is $\lambda\eq 3/2$ 
(the actual value is not that important as long as it is small enough to lead to black hole 
formation instead of dispersion which is the case here). It is convenient to make the change of variable $ r \rightarrow x = \frac{r}{r + 2}$. This reduces the overall number of grid points while maintaining a high density of points in the region where large gradients are present at late times. The numerical code is based on a fourth order Adams-Bashforth-Moulton explicit scheme \cite{Butcher, Press} with space and time increments equal to $\Delta x \eq\Delta t \eq 1\times 10^{-4}$. The total energy $E_{tot}$ is calculated from the integral form of the ADM mass \reff{4DScalarEtotInt}. As already mentioned, the mass $E$ of the black hole will be less than the total energy $E_{tot}$ because some energy appears in an outgoing matter wave. The ADM mass $E_{tot}$ is a conserved quantity which should remain constant during the evolution and we therefore use it to monitor the accuracy of the simulation. We show the plots up to $t\eq 20$ at which time the function $F$ has plateaued and $E_{tot}$ is just beginning to deviate from unity. 

\begin{figure}[tbp]
		\begin{center}
			\includegraphics[scale=.38, draft=false, trim=2cm 1.5cm 2.5cm 2cm, clip=true]{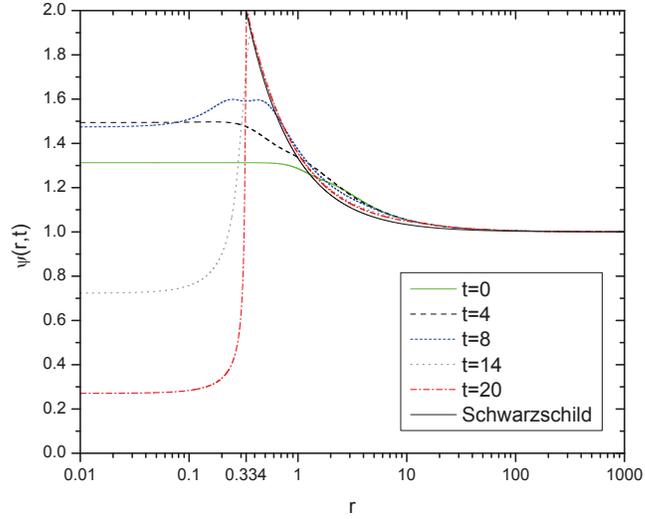}
		\end{center}
		\caption{\label{4DScalarGraPsi} Space profile of the metric field $\psi$ for different times.  At late times ($t \eq 20$), the exterior region matches almost exactly the Schwarzschild form.}
	\end{figure}

\begin{figure}[tbp]
		\begin{center}
			\includegraphics[scale=0.38, draft=false, trim=2cm 1.5cm 2.5cm 2cm, clip=true]{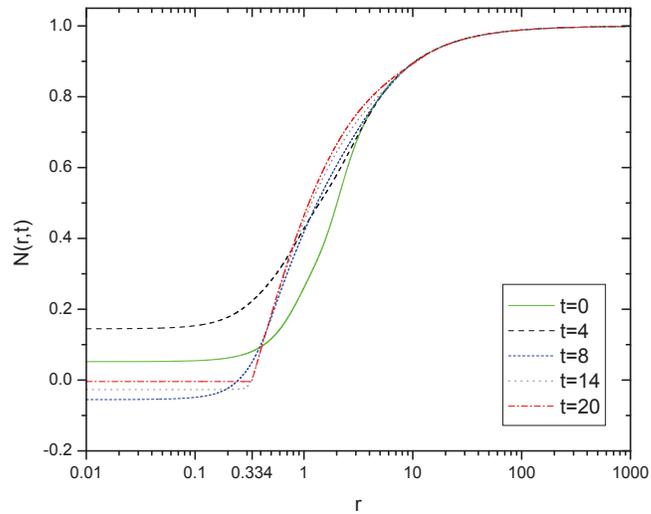}
		\end{center}
		\caption{\label{4DScalarGraN} Space profile of the lapse function $N$ for different times. The event horizon ($r \eq 0.334$) is the radius where $N$ crosses zero at late times. After crossing zero at $r\eq 0.334$, $N$ is negative in the interior and approaches zero from below.}
	\end{figure}

\begin{figure}[tbp]
		\begin{center}
			\includegraphics[scale=.37, draft=false, trim=2cm 1.5cm 2.5cm 2cm, clip=true]{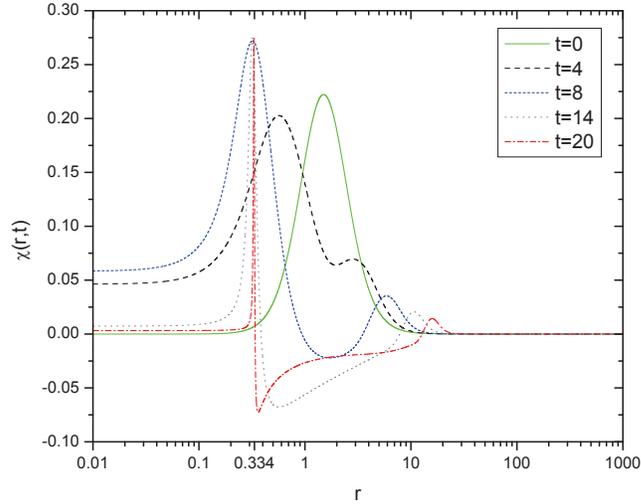}
		\end{center}
		\caption{\label{4DScalarGraChi} Space profile of the matter field $\chi$ for different times. The initial matter distribution is a shell. It first expands into the interior ($t\eq 4$) before collapsing to a thin shell near the event horizon at late times. Note the propagation of an outgoing matter wave in the exterior region.}
	\end{figure}

The metric functions $\psi(r,t)$ and $N(r,t)$ as well as the matter function $\chi(r,t)$  are plotted as functions of $r$ (on a log scale) at different times $t$ in figures \ref{4DScalarGraPsi}, \ref{4DScalarGraN} and \ref{4DScalarGraChi} respectively. At late times, in the exterior region, the conformal factor $\psi$ (fig. \ref{4DScalarGraPsi}) can be seen to match closely the analytical form $\psi=1+ M/2r$ of the Schwarzschild metric \reff{isotropic}. This provides a check on the numerical simulation and confirms that the collapse process has indeed led to a Schwarzschild black hole. The event horizon is at $r\eq 0.334$, the location where the lapse function $N(r,t)$ (fig. \ref{4DScalarGraN}) crosses zero at late times.  After $N$ crosses zero, it is negative throughout the interior region. Inside, it approaches zero from below but never actually crosses zero. The spacetime is well behaved throughout the evolution and all curvature scalars are finite. From the plot of the matter field $\chi$ (fig. \ref{4DScalarGraChi}), one observes that at $t\eq 4$, the initial shell configuration has expanded into the interior to form a spherical ball of matter. By that time, it has not yet collapsed to a black hole since $N$ has not crossed zero yet ( $N$ crosses zero for the first time at around $t\eq 6$). At late times, the scalar field $\chi$ has collapsed to a thin shell \cite{Edery_Hugues} near the event horizon except for an outgoing matter wave in the exterior region. The crest of the wave at $t\eq 8$ and $r\approx 5$ propagates outwards and reaches $r \approx 10$ at $t\eq14$ and $r\approx 15$ at $t\eq20$. 
	
\begin{figure}[tbp]
		\begin{center}
			\includegraphics[scale=.38, draft=false, trim=2cm 1.5cm 2.5cm 2cm, clip=true]{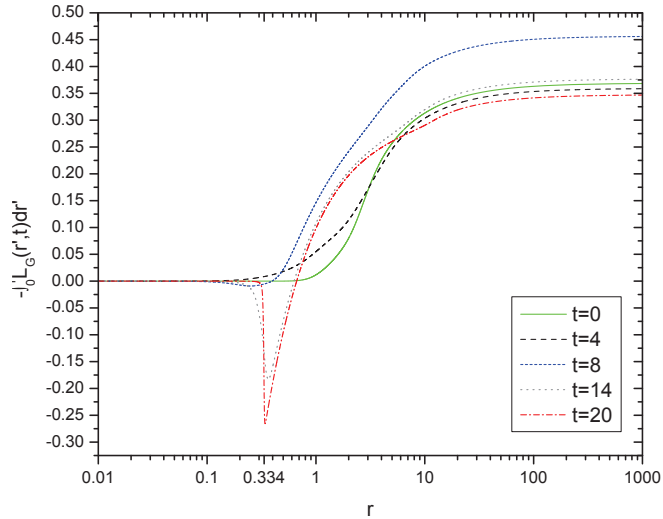}
		\end{center}
		\caption{\label{4DScalarGraCumulGrav} Accumulation plot for the gravitational Lagrangian at different times. At late times, there is a negative contribution (the large dip) stemming from a thin slice just inside the event horizon where the metric field $\psi$ is nonstationary.}
	\end{figure}

\begin{figure}[tbp]
		\begin{center}
			\includegraphics[scale=.38, draft=false, trim=2cm 1.5cm 2.5cm 2cm, clip=true]{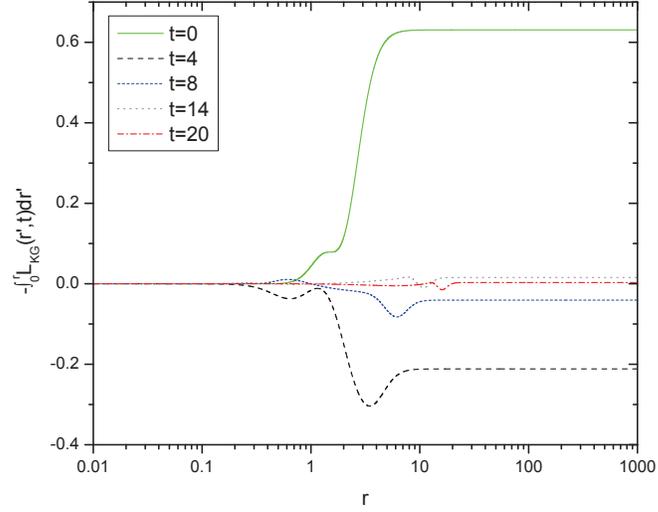}
		\end{center}
		\caption{\label{4DScalarGraCumulMatt} Accumulation plots for the Klein-Gordon Lagrangian for different times. It is close to zero at late times.}
	\end{figure}

\begin{figure}[tbp]
		\begin{center}
			\includegraphics[scale=.38, draft=false, trim=2cm 1.5cm 2.5cm 2cm, clip=true]{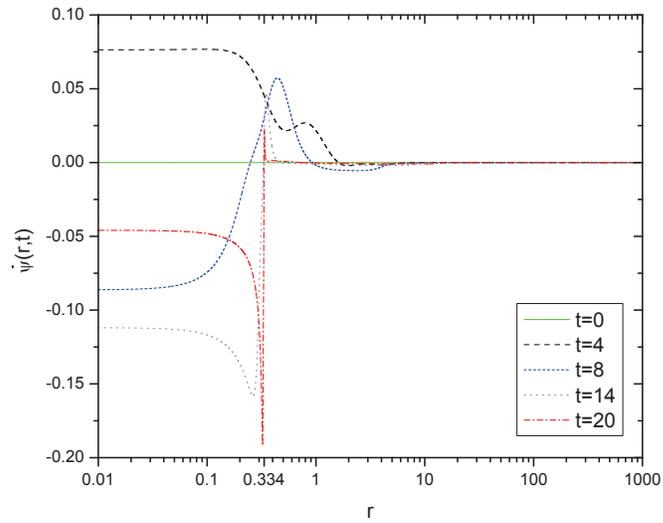}
		\end{center}
		\caption{\label{4DScalarGraPsiDot} Time dependence of the metric field $\psi$. $\dot{\psi}$ peaks just inside the event horizon, is decreasing in magnitude in the rest of the interior and is zero outside.}
	\end{figure}

In the function $F\eq -L$, there are two contributions to the total Lagrangian $L$ given by \reff{4DScalarLtot}: one from gravitation ($L_G$) and one from matter ($L_{\kg}$). Both $L_G$ and $L_{\kg}$ are expressed as integrals over $r$ where $r$ ranges from $0$ to the outer computational boundary $R$. It is instructive to see where $F$ accumulates as the integral ranges from $0$ to $R$. These accumulation plots are shown at different times $t$ for gravitation 
(fig. \ref{4DScalarGraCumulGrav}) and matter (fig. \ref{4DScalarGraCumulMatt}). At late stages of the collapse, the matter contribution to the free energy is almost zero: the free energy stems almost entirely from the gravitational part and we see that black hole entropy is {\it gravitational} entropy. In fig. \ref{4DScalarGraCumulGrav}, at the late time $t \eq 20$, there is a dip in a thin region $\Re$ of thickness $\Delta r=\epsilon$ just inside the event horizon $(0.334-\epsilon \less r \less 0.334)$. The region $\Re$ is nonstationary: as can be seen from fig. \ref{4DScalarGraPsiDot},  $\dot{\psi}$ peaks in $\Re$, is zero in the exterior region and tends with time towards smaller values in the interior region outside $\Re$. In short, $\psi$ is basically static everywhere except in the nonstationary region $\Re$. In the previous section we saw that the nonstationary interior makes a negative contribution to the free energy of a Schwarzschild black hole whereas the static exterior makes a positive contribution. Our numerical plots bear this out. The change in free energy in the interior region, the dip in fig. \ref{4DScalarGraCumulGrav}, is negative and this occurs in the nonstationary region $\Re$ whereas in the exterior, which is basically static except for an outgoing matter wave, the change in free energy in fig. \ref{4DScalarGraCumulGrav} is positive. We now compare numerical values of $F\eq -L$ during gravitational collapse to the free energy of a Schwarzschild black hole.                

\begin{figure}[tbp]
		\begin{center}
			\includegraphics[scale=.38, draft=false, trim=2cm 1.5cm 2.5cm 2cm, clip=true]{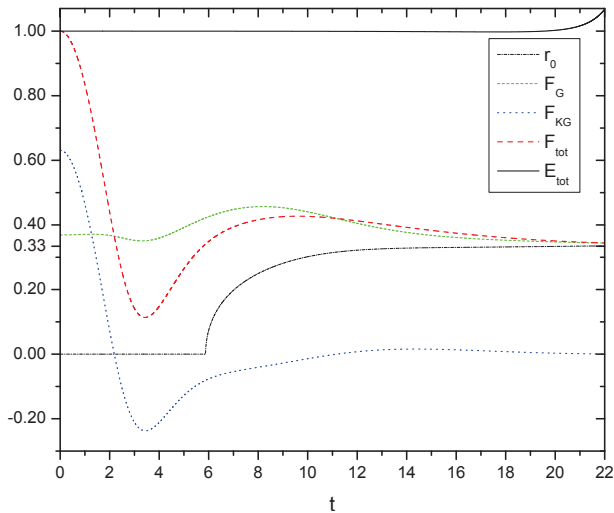}
		\end{center}
		\caption{\label{4DScalarGraAct} Time profile of the free energy and the horizon radius. The total free energy $F_{tot}$ and the horizon radius $r_0=E/2$ converge towards the same value at late times.}
	\end{figure}

Figure \ref{4DScalarGraAct} shows the time evolution of the function $F\eq -L$, along with its gravity and matter contribution.  At the late time $t\eq 20$, $F$ has a numerical value of $0.351$. This needs to be compared to $E/2$, the value of the free energy of a $4D$ Schwarzschild black hole of mass $E$. The value of $E/2$ is obtained directly from the horizon radius $r_0=E/2$, the location where the lapse function $N$ crosses $0$ at late times. Note how the plots of $r_0$ and $F_{tot}$ merge together at late times in fig. \ref{4DScalarGraAct}. The value of $r_0$ at $t\eq 20$ is $0.334$. The value $F\eq 0.351$ obtained numerically from $-L$ is therefore equal to the $E/2$ result ($0.334$) to within $5\%$. The 5\%  discrepancy is a conservative value. It is based on a  comparison between  the free energy $F$ and $E/2$ at t=20 where $E_{tot}$ was just starting to deviate from unity. From t=20 to t=22, the numerical curve of the metric function $\psi$ in most of the exterior continues to approach closer to the analytical curve $\psi=1+ M/2r$. This suggests that the evolution from t=20 to t=22 is still good even though $E_{tot}$ has deviated from unity. In Fig.7, the free energy F and the E/2 curve continue to approach each other in the time interval from t=20 to t=22 and the discrepancy at t=22 is 2.7\% (F=0.344 and E/2=0.335)\footnote{The matter contribution to the free energy from t=20 to t=22 is separately evolving towards its expected theoretical value of zero: it is 0.003 at t=20 and 0.0004 at t=22.}. This can be viewed as a more realistic value for the percentage discrepancy.  It is clear from this analysis that the main reason for the 5\% discrepancy is that F and E/2 have not evolved far enough in time to reach a plateau before $E_{tot}$ starts to deviate from unity. The goal in the future would then be to push the time further numerically. It should be possible to accomplish this by increasing the resolution and implementing adaptive mesh refinement \cite{Oliger} to reduce the computing time (we ran the code with the larger step size of $\Delta x=0.001$, $\Delta t=0.001$ and the same $\lambda=1.5$. At t=20, the discrepancy is $14\%$ compared to $5\%$ and the $E_{tot}$ is quite off so that the higher resolution of $\Delta x=0.0001$, $\Delta t=0.0001$ leads clearly to better results). It is worth noting that the discrepancy also depends on the initial state. Recall that the initial field configuration for the scalar field $\chi$ depends on the scale parameter $\lambda$ \reff{4DScalarIEChi}. We ran the code for different values of $\lambda$ (albeit at at the lower resolution of $\Delta x=0.001$, $\Delta t=0.001$ to speed up the simulation). We observed lower discrepancies for lower $\lambda$ in a given range ($30\%$ at $\lambda=1.55$, $12\%$ at $\lambda=1.5$ and $10\%$ at $\lambda=1.49$). The value of $\lambda$ also affects the duration of the simulation, extending the time where $E_{tot}$ is close to unity from T=15 for $\lambda = 1.55$ up to T=17.5 for $\lambda = 1.49$.          

\clearpage
	
\clearpage
		
	\clearpage
	
\clearpage

\section{Free energy during the collapse of a 5D massless scalar field} 

Though $3\p 1$ dimensions represents the most realistic scenario, it would be short sighted not to consider the higher five-dimensional (4+1) case. Besides providing another arena where the thermodynamics of gravitational collapse can be studied numerically, $5D$ gravitational collapse has had surprising connections to $4D$ physics. For example, a holographic argument backed up with numerical results was offered for a connection between the Choptuik critical scaling exponent, $\gamma_{5D}$, in 5-dimensional black hole formation via scalar field collapse \cite{Sorkin,Bland} and the saturation exponent, $\gamma_{BFKL}$, of four-dimensional Yang-Mills theory in the Regge limit \cite{Alvarez}. The latest value of $\gamma_{5D} = 0.4131\pm 0.0001$  obtained from numerical work on $5D$ gravitational collapse \cite{Kunstatter} was close to the value of $\gamma_{BFKL} = 0.409552$  obtained from numerical work on $4D$ Yang-Mills \cite{Alvarez}. The near matching of the two results provides support for the conjecture relating $5D$ gravitational physics to pomeron exchange in $4D$ Yang-Mills theory \cite{Alvarez}. We have learned by now that gravitational investigations in five dimensions are not just a curiosity but have something real to say about our four dimensional world.  

In this section we track the free energy $F\eq-L$ during the classical gravitational collapse of a massless scalar field in five dimensional isotropic coordinates. Whereas the previously studied $5D$ Yang-Mills instanton \cite{Khlebnikov} acts like dust and collapses to a black hole for any initial static configuration, a static massless scalar field has pressure and two outcomes are possible:  dispersion or gravitational collapse to a black hole \cite{Choptuik}. The initial configuration of the scalar field is governed by a scale parameter $\lambda$ and the system collapses to a black hole for sufficiently small values of $\lambda$. At late stages of the collapse process, we obtain a numerical value of the function $F=-L$ that is within $5\%$ of the free energy of $E/3$ obtained from $5D$ standard black hole thermodynamics. Although a massless scalar field and a Yang-Mills instanton constitute different kinds of matter, they collapse to nearly the same final thermal state. We see in a dynamical fashion that black hole thermodynamics is not dependent on the type of matter undergoing the collapse.

A spherically symmetric time-dependent $5D$ metric in isotropic coordinates takes the form 
\beq
	ds^2 = -N(r,t)^2 dt^2 + \psi(r,t)^2 (dr^2 + r^2 d\Omega_3^2),
\eeq{isometric5D}
where $d\Omega_3^2 = d\theta^2 + \sin^2\theta ( d\phi^2 + \sin^2\phi\, d\gamma^2)$ is the metric on $S^3$, the unit 3-sphere, with  $0\le\theta\le\pi$, $0\le\phi\le\pi$ and $0\le\gamma\le 2\,\pi$. As before, $N(r,t)$ is called the lapse function and $\psi(r,t)$ the conformal factor. In $5D$ isotropic coordinates, the {\it exterior} region of the Schwarzschild metric is given by \cite{Khlebnikov}
\beq
	ds^2 = - \dfrac{\big(1-\frac{r_0^2}{r^2}\big)^2}{\big(1+\frac{r_0^2}{r^2}\big)^2}dt^2 + \Big(1+\frac{r_0^2}{r^2}\Big)^2 (dr^2 + r^2 d\Omega_3^2),
\eeq{isotropic5D}
where $r_0$ is the location of the event horizon and is given by $r_0^2 = 2\,G_5 \,E/(3\pi)$ ($G_5$ is Newton's constant in $5D$ and $E$ is the ADM mass of the black hole). The static metric \reff{isotropic5D} does not describe the interior of the $5D$ Schwarzschild black hole, which is nonstationary. During the collapse process, the metric \reff{isometric5D} approaches the metric \reff{isotropic5D} only in the exterior region outside the event horizon. In the interior, the lapse function $N(r,t)$ and conformal factor $\psi(r,t)$ deviate from \reff{isotropic5D} and are time-dependent in accordance with the nonstationary nature of the interior.    

\subsection{Gravitational and matter Lagrangian in 5D}

The Einstein-Hilbert action in $5D$ is given by 
		\begin{equation}
			I_{\eh} = \frac{1}{16\pi \, G_5} \int \sqrt{-g_5}\,R_5 d^5x
		\end{equation}
		where the Ricci scalar $R_5$ evaluated for the isotropic metric \reff{isometric5D} is
		\begin{eqnarray}
			R_5 &=& \frac{2}{r\psi^3N^3} ( 4r\ddot{\psi}\psi^2N - 4r\dot{N}\dot{\psi}\psi^2 - 9\psi'N^3 - 3r\psi''N^3 \\
			& & {} - 2rN'\psi'N^2 - 3N'N^2\psi + 6r\dot{\psi}^2N\psi - rN''N^2\psi ), \nonumber
		\end{eqnarray}
		and $\sqrt{-g_5} \eq N\psi^4\,r^3\sin^2\theta\sin\phi$. Integrating by parts the second derivatives $N''$, $\psi''$, and $\ddot{\psi}$ we obtain
		\begin{equation}
			I_{\eh} = \int L_{G} \,dt - I_B,
		\end{equation}
	where
		\begin{equation}
			L_{G} = \frac{\pi}{8\,G_5} \int \Big( -\frac{12r^3\dot{\psi}^2\psi^2}{N} + 6r^3N'\psi\psi' + 6r^3N\psi'^2 \Big) \,dr
		\end{equation}
is the gravitational Lagrangian and $I_B$ is a boundary term \cite{Poisson}. Variation with respect to the metric of the gravitational action $I_G \eq I_{\eh} + I_B \eq \int L_{G} \,dt$ reproduces the Einstein field equations. 
	
The Klein-Gordon action for a massless scalar field in $5D$ is
\beq
			I_{\kg} = -\frac{1}{2} \int \sqrt{-g_5}\, g^{\mu\nu} \partial_{\mu}\chi \, \partial_{\nu}\chi \, d^5x = \int \sqrt{-g_5} \,\mathcal{L}_{\kg} \,d^5x
\eeq{} where spherical symmetry implies $\chi\equiv\chi(r,t)$ and $\mathcal{L}_{\kg} \eq (-1/2)\,g^{\mu\nu}\partial_{\mu}\chi\,\partial_{\nu}\chi$ is the Lagrangian density. The matter Lagrangian evaluated for the isotropic metric \reff{isometric5D} is given by, 
		\begin{equation}
			L_{\kg} = \int \sqrt{-g_5} \,\mathcal{L}_{\kg} \,d^4x = -\pi^2 \int N\,\psi^4\,r^3  \Big(\frac{\chi'^2}{\psi^2} - \frac{\dot{\chi}^2}{N^2}\Big) \,dr \,.
		\end{equation}
	
The total Lagrangian $L_{tot}=L_G+L_{\kg}$ is then
		\beqa
			L_{tot} &=& \frac{\pi}{8\,G_5} \int \Big( -\frac{12r^3\dot{\psi}^2\psi^2}{N} + 6r^3N'\psi\psi' + 6r^3N\psi'^2 \Big) \,dr \\
			& & {} - \pi^2 \int N\psi^4r^3 \Big( \frac{\chi'^2}{\psi^2} - \frac{\dot{\chi}^2}{N^2} \Big) \,dr \,. \nonumber
		\eeqa{5DScalarLtot}
During the collapse process we track the function $F(t)\eq -L_{tot}$. 
	
\subsection{5D equations of motion}

In $4D$, the Einstein field equations yield the energy constraint \reff{4DScalarEConstraint}, the momentum constraint \reff{4DScalarMConstraintA}, the evolution equation \reff{4DScalarEqPsi} for the conformal factor $\psi$, the evolution equation \reff{4DScalarEqK} for its ``momentum conjugate", the trace of the extrinsic curvature $K$ and finally the ODE \reff{4DScalarEqN} for the lapse function $N$. The corresponding equations in $5D$ are (see derivation in Appendix A)
\begin{align}
&\word{energy constraint:} 	-3\frac{\nabla^2\psi}{\psi^3} = \kappa^2\mathcal{E} - \frac{3}{8}K^2 \label{5DScalarEConstraint}\\
&\word{momentum constraint:}	\frac{K'}{4} = \frac{\kappa^2}{3}\frac{\dot{\chi}}{N}\chi'\label{5DScalarMConstraintA}\\
&\word{evolution equation for} \psi: 	\frac{\dot{\psi}}{N} = -\frac{K\psi}{4}\label{5DScalarEqPsi}\\
&\word{evolution equation for K:} \frac{\dot{K}}{N} = \frac{K^2}{2} (1+w) + \frac{4}{3}\kappa^2\mathcal{E} (1-w)	\label{5DScalarEEqK}\nonumber\\
&\qquad\qquad\qquad\qquad\qquad\qquad\qquad\qquad- 4\frac{\psi'}{\psi^3} \Big( \frac{\psi'}{\psi} + \frac{2}{r} \Big) - 4\frac{N'}{N\psi^2} \Big( \frac{\psi'}{\psi} + \frac{1}{r} \Big) - 4w\frac{\nabla^2\psi}{\psi^3}\\
&\word{ODE for N:} \frac{2r}{\psi}\partial_r \Big( \frac{\psi'}{r\psi^2} \Big) + \frac{r}{N}\partial_r \Big( \frac{N'}{r\psi^2} \Big) = -\kappa^2 \frac{\chi'^2}{\psi^2} 	\label{5DScalarEqN}
\end{align}
where the arbitrary parameter $w$ in \reff{5DScalarEEqK} arises from a term added for numerical stability purposes and $\mathcal{E}$ is interpreted as the energy density of the matter field and is given by
\begin{equation}
			\mathcal{E}= \frac{1}{2} \Big( \frac{\chi'^2}{\psi^2} + \frac{\dot{\chi}^2}{N^2} \Big).
		\label{5DScalarEDensity}
		\end{equation}
Note that in $5D$ the flat space Laplacian is given by 
\begin{equation}
			\label{5DScalarLaplacian}
			\nabla^2\psi = \frac{1}{r^3}\partial_r ( \psi'r^3 ) = \frac{3}{r}\psi' + \psi'' \,.
		\end{equation}
There are also the evolution equations for the matter field $\chi$ and its ``momentum conjugate" $p$. These are  
\beq
\frac{\dot{\chi}}{N} = \frac{p}{\psi^4}
\eeq{5DScalarEqChi}
and
	\beq
		\frac{\dot{p}}{N} = \psi^2 \Big( \chi'' + \frac{N'}{N}\chi' + 2\frac{\psi'}{\psi}\chi' + \frac{3}{r}\chi'\Big)
	\eeq{5DScalarEqP}

\subsection{Initial states}

The initial states for the matter and metric fields are chosen to be static: $K \eq p \eq 0$ or $\dot{\psi} \eq \dot{\chi} \eq 0$ at $t \eq 0$. The initial configuration of the massless scalar field $\chi$ is chosen to be a shell with Maxwellian distribution:
		\beq
			\chi(r,t=0) = \frac{1}{\lambda^3}r^2e^{\frac{-r^2}{2\lambda^2}}
		\eeq{5DScalarInitialChi}
where the scale parameter $\lambda$ is a positive constant. The function peaks at $r \eq \sqrt{2}\,\lambda$ and the half-width of the shell is $\approx 1.6\lambda$. The function is zero at the two extremities: $\chi(r,t\eq0) \to 0$ as $r \rightarrow 0$ and as $r \rightarrow \infty$.  A large $\lambda$ implies a diluted energy density and a weaker self-gravitational force. As in 4D, black hole formation occurs only for a sufficiently small value of $\lambda$. 

The initial state for the conformal factor $\psi$ is obtained by substituting the above initial profile for $\chi$ into the energy constraint \reff{5DScalarEConstraint} and solving it order by order in powers of $\zeta=\kappa^2$ as we did for the $4D$ case (we refer the reader to section 2.3 for the procedure and we will simply state the result here).  The initial state for $\psi$ has to be expanded to at least order $\zeta^3$ for the simulation to be sufficiently accurate. Its expression is
\beqa 
	\psi(r,t=0) \!\!\!&\eq&\!\! 1 + \frac{\zeta}{24\,r^2\lambda^6}\,e^{\frac{-r^2}{\lambda^2}} \Big[ -r^6 - 2\,r^4\lambda^2 - 6\,r^2\lambda^4 +  8\lambda^6 \big( e^{\frac{r^2}{\lambda^2}} -1 \big) \Big] \nonumber\\
	& &\!\!\!\!\! {} + \frac{\zeta^2}{4608\,r^2\lambda^{12}}\,e^{\frac{-2r^2}{\lambda^2}} \Big[2\,r^{10} + 6\,r^8\lambda^2 + \ldots - 3\,r^2\lambda^8 + \lambda^{10} \big( 77e^{\frac{2r^2}{\lambda^2}} - 128e^{\frac{r^2}{\lambda^2}} + 51 \big) \Big] \nonumber\\
	& &\!\!\!\!\! {} + \frac{\zeta^3}{241864704\,r^2\lambda^{18}}\,e^{\frac{-3r^2}{\lambda^2}} \Big[ 486\,r^{14} + 1782\,r^{12}\lambda^2 + \ldots + 6\,r^2\lambda^{12} \big( 11664\,e^{\frac{r^2}{\lambda^2}} - 2869 \big) \nonumber\\
	& &\!\!\!\!\! {} \quad\quad- 2\,\lambda^{14} \big( 105832\,e^{\frac{3r^2}{\lambda^2}} - 168399e^{\frac{2r^2}{\lambda^2}} + 78732\,e^{\frac{r^2}{\lambda^2}} - 16165 \big) \Big]. \nonumber
\eeqa{5DScalarExpansionPsi}
The initial state for $N$ is obtained numerically from its ODE \reff{5DScalarEqN}, using the initial states for $\psi$ and $\chi$ and iterating backwards starting with the asymptotic value $N(r=R,t)\eq 1$.

\subsection{ADM mass in 5D}

To obtain an expression for the ADM mass in $5D$, one uses the same definition \reff{ADMmass} but now integrated over $d^3\theta$,
\beq
	M_{ADM} = -\frac{1}{8\pi} \lim_{S_t \rightarrow \infty} \oint_{S_t} (k-k_0) \sqrt{\sigma} \, d^3\theta,
\eeq{ADMmass5D}
where $S_t$ is the three-sphere boundary. Let $\Sigma_t$ be the spacelike hypersurface at constant time. For the metric \reff{isometric5D}, the trace of the extrinsic curvature of $S_t$ embedded in $\Sigma_t$ is given by $k = \nabla_a r^a = 3\psi' / \psi^2 + 3/(\psi \,r)$ where $r_a = \psi\partial_a r$ is the unit normal vector to the three-sphere boundary $S_t$, while the trace of the extrinsic curvature of $S_t$ embedded in flat spacetime is $k_0 = 3/(\psi \,r)$. With $\sqrt{\sigma}\, d^3\theta = \psi^3\,r^3\,\sin^2\theta\,\sin\phi\,d\theta\,d\phi\,d\gamma$, the ADM mass for the $5D$ isotropic metric \reff{isometric5D} reduces to
\beq
	M_{ADM} = -\dfrac{3\pi\, r^3}{4 G_5} \psi\,\partial_r\psi\big|_{r=R}= -\dfrac{3\pi R^3}{4 G_5} \partial_r\psi\big|_{r=R}\,\,\,
\eeq{5DScalarADMMass}
where $\psi=1$ at $r=R$ was used. Using the energy constraint equation \reff{5DScalarEConstraint}, one can express the ADM mass in the integral form
	\beq
		M_{ADM} = E_{tot} = 2\,\pi^2\,\int^R_0 \Big(\mathcal{E} - \dfrac{3K^2}{8\kappa^2} \Big)\psi^3r^3\,dr\,.
	\eeq{5DScalarEtotInt}
The ADM mass represents the total energy $E_{tot}$. As in $4D$, the black hole mass $E$ is less than $E_{tot}$ because part of the total energy is carried away by an outgoing matter wave. In $5D$, $E$ is obtained via the relation $r_0^2=2\,G_5\,E/(3\,\pi)$ where the gravitational radius $r_0$ is the location where the lapse function $N$ crosses zero at late times. 

\subsection{5D results}

We work in the following geometrized units: $G_5\eq 1/(6\pi)$ and $c\eq1$. In 5D geometrized units, mass has dimension of length squared (this is why in 5D, the black hole mass $E$ is proportional to the square of gravitational radius $r_0$). Masses will be expressed in terms of the ADM mass. The ADM mass depends on the parameter $\lambda$ which sets the length scale. We use the same numerical technique as in 4D (a fourth-order ABM scheme). We perform the spatial coordinate change $r \rightarrow x = \frac{r}{r + 2}$, which allows for more grid points near $r\eq1$ where the gradients are largest. The space and time intervals used are $\Delta x = 1 \times 10^{-4}$ and $\Delta t = 5 \times 10^{-5}$ respectively. The value of the scale parameter $\lambda$ has to be small enough for gravitational collapse to occur and our simulation is run using $\lambda \eq 1$ which leads to black hole formation instead of dispersion. The value of $\lambda$ affects the time of collapse but not the black hole thermodynamics at late times.

\begin{figure}[tbp]
		\begin{center}
			\includegraphics[scale=.38, draft=false, trim=2cm 1.5cm 2.5cm 2cm, clip=true]{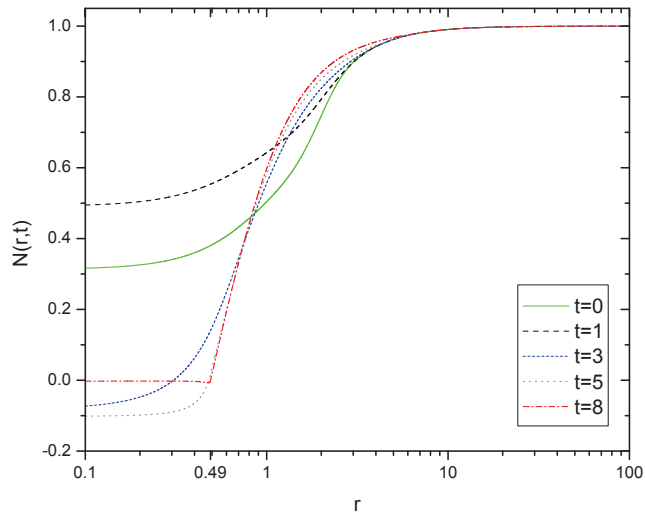}
		\end{center}
		\caption{\label{5DScalarGraN} Space profile of the lapse function $N$ at different times. $N$ crosses zero at the location of the apparent horizon. This occurs for the first time at $t\eq 2.6$. At late times, $N$ crosses zero at $r \eq 0.49$ and this is identified as the radius of the event horizon.  After $N$ crosses zero at $r\eq 0.49$, it is negative (albeit very small) throughout the interior ($r<0.49$) and never crosses zero inside.}  
	\end{figure}

\begin{figure}[tbp]
		\begin{center}
			\includegraphics[scale=.38, draft=false, trim=2cm 1.5cm 2.5cm 2cm, clip=true]{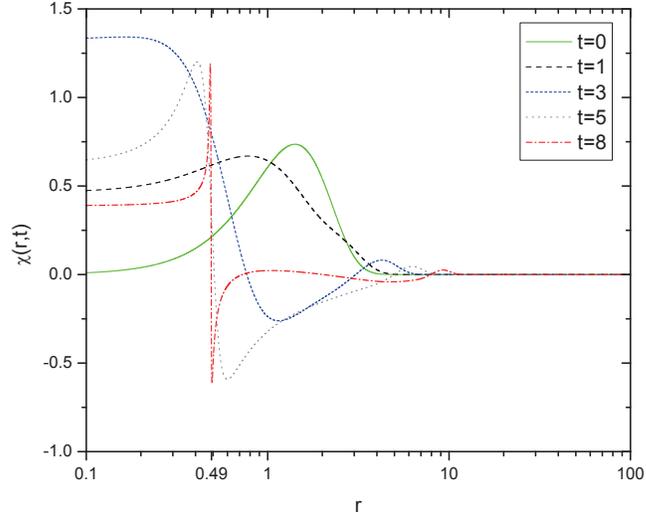}
		\end{center}
		\caption{\label{5DScalarGraChi} Space profile of the matter field $\chi$ at different times. At late times, the matter has collapsed to a thin shell near the event horizon. Note the propagation of an outgoing matter wave in the exterior region.}
	\end{figure}

\begin{figure}[tbp]
		\begin{center}
			\includegraphics[scale=.38, draft=false, trim=2cm 1.5cm 2.5cm 2cm, clip=true]{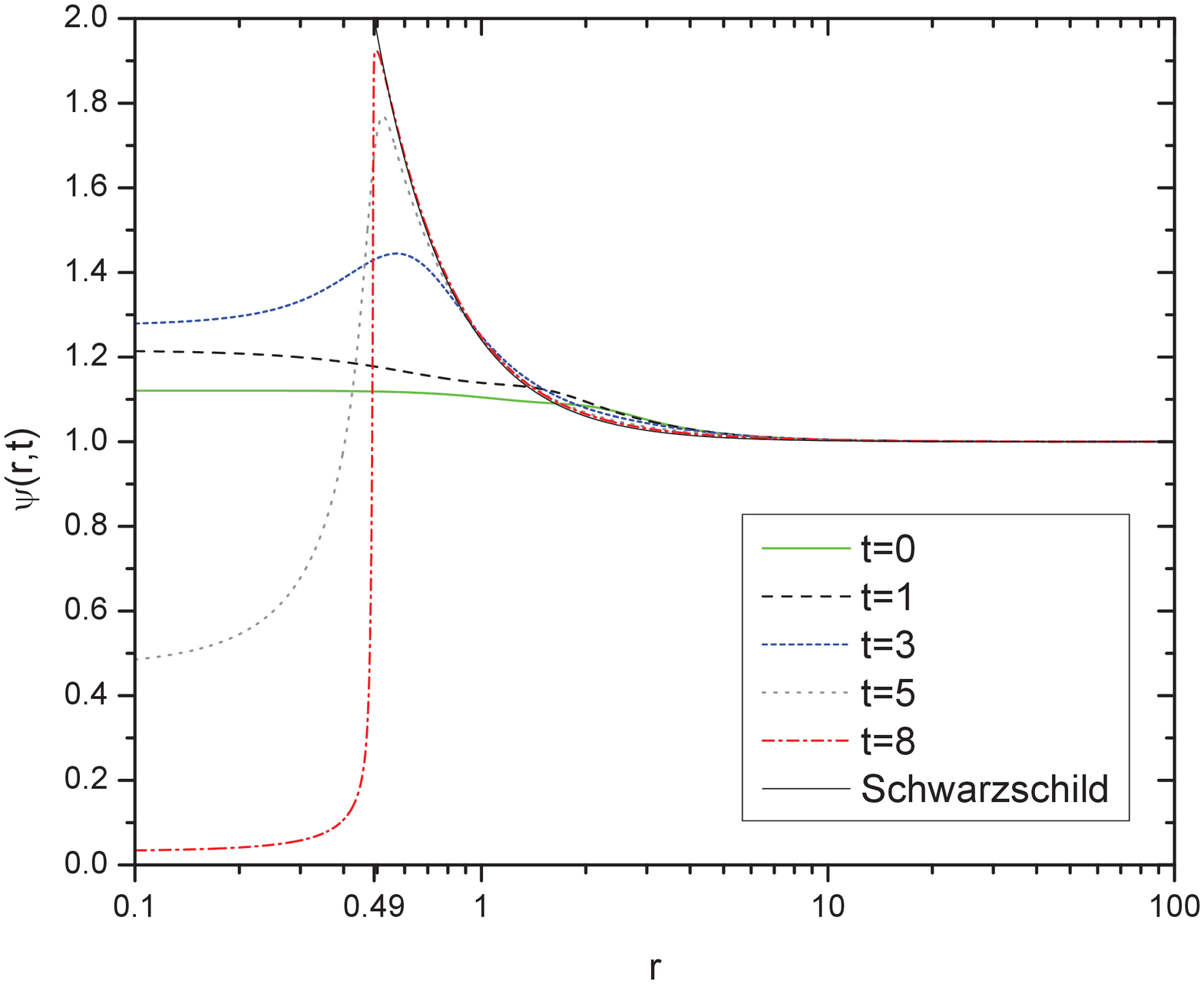}
		\end{center}
		\caption{\label{5DScalarGraPsi} Space profile of the metric field $\psi$ at different times.  At late times ($t \eq 8$), the exterior region matches almost exactly the Schwarzschild form. Note that 
		$\psi\to 0$ in the interior.}
	\end{figure}

Black hole formation occurs when the lapse function $N$ in fig. \ref{5DScalarGraN} first crosses zero. The data shows that this occurs for the first time around $t\eq 2.6$. At late stages of the collapse, $N$ crosses zero at $r\eq 0.49$, which is associated with the event horizon. In the interior region $(r<0.49)$, $N$ becomes negative and tends to (but never reaches) zero. The spacetime is well behaved throughout the simulation. Figure \ref{5DScalarGraChi} shows the time evolution of the matter field $\chi$.  Early in the simulation, the shell expands into a spherical ball, filling the interior, before collapsing to a black hole at around $t=2.6$.  At late times, the matter field has collapsed to a thin shell immediately inside and on the horizon, except for an outgoing matter wave located at around $r=10$ at $t=8$. In the region outside the event horizon, the conformal factor $\psi$ (fig. \ref{5DScalarGraPsi}) matches closely the analytical Schwarzschild form given by metric \reff{isometric5D} so that the collapse process has indeed led to a Schwarzschild black hole. 

\begin{figure}[tbp]
		\begin{center}
			\includegraphics[scale=.38, draft=false, trim=2cm 1.5cm 2.5cm 2cm, clip=true]{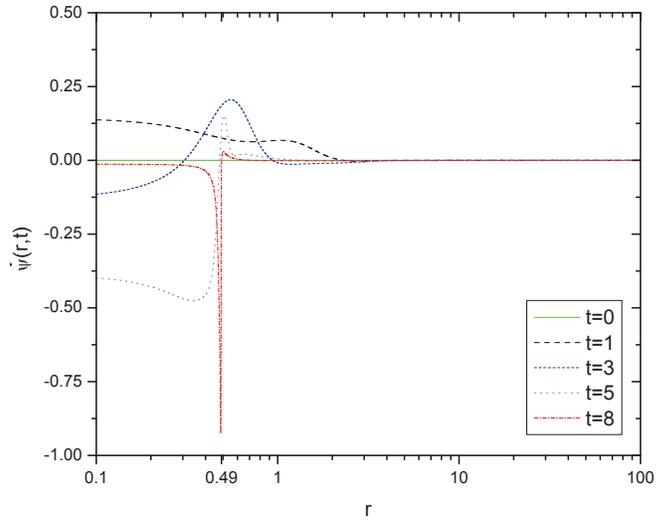}
		\end{center}
		\caption{\label{5DScalarGraPsiDot} Time dependence of the metric field $\psi$. It is static except for a thin slice just inside the event horizon.}
	\end{figure}

\begin{figure}[tbp]
		\begin{center}
			\includegraphics[scale=.38, draft=false, trim=2cm 1.5cm 2.5cm 2cm, clip=true]{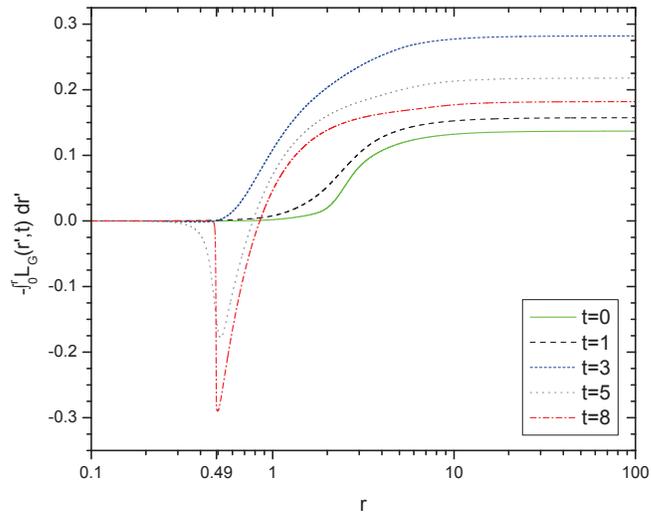}
		\end{center}
		\caption{\label{5DScalarGraCumulGrav} Accumulation plot for the gravitational Lagrangian at different times. There is a negative contribution (the large dip) coming from the region just inside the event horizon.}
	\end{figure}

\begin{figure}[tbp]
		\begin{center}
			\includegraphics[scale=.38, draft=false, trim=2cm 1.5cm 2.5cm 2cm, clip=true]{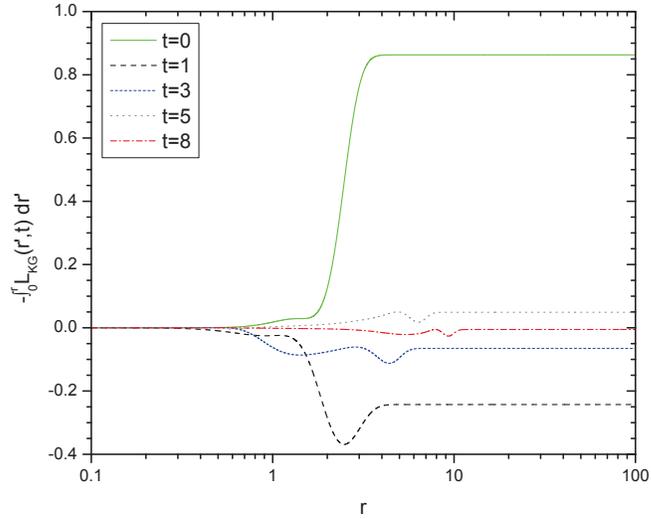}
		\end{center}
		\caption{\label{5DScalarGraCumulMatt} Accumulation plot for the Klein-Gordon Lagrangian at different times. It approaches zero at late times except for the small contribution of the outgoing matter wave.}
	\end{figure}

\begin{figure}[tbp]
		\begin{center}
			\includegraphics[scale=.38, draft=false, trim=2cm 1.5cm 2.5cm 2cm, clip=true]{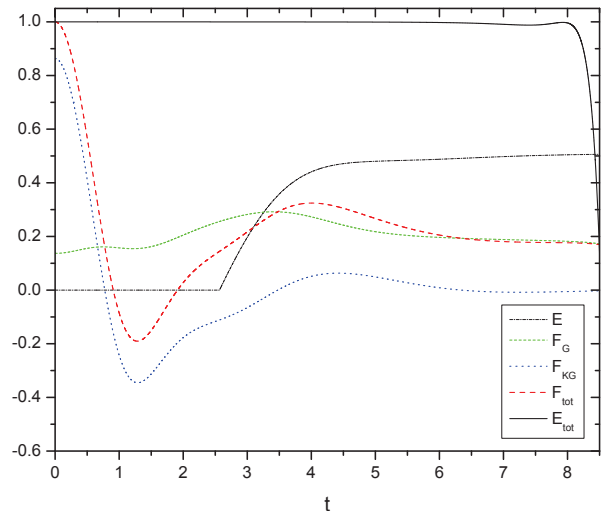}
		\end{center}
		\caption{\label{5DScalarGraAct} Time profile of the free energy and the mass of the black hole. The total free energy approaches a value of a third of the mass $E$ at late times.}
	\end{figure}

The temporal dependence of the metric field $\psi$ is shown in fig. \ref{5DScalarGraPsiDot}. At late times, the metric field $\psi$ is nonstationary in a small region just inside the horizon. This reflects the fact that the Schwarzschild spacetime is nonstationary in the interior region. It is this small nonstationary region which is responsible for the large dip near the event horizon ($r \eq 0.49$) at late times in the accumulation plot of the free energy for gravity (see fig. \ref{5DScalarGraCumulGrav}). In accordance with our discussion in section 2.4.1, the nonstationary interior region makes a negative contribution to the free energy, whereas the basically static exterior makes a positive contribution. The matter contribution to the free energy (see fig. \ref{5DScalarGraCumulMatt}) is essentially zero except for a small contribution from the outgoing matter wave. The free energy is almost entirely gravitational in origin and in  fig. \ref{5DScalarGraAct}, the total free energy $F_{tot}$ is equal to the gravitational contribution $F_G$ at late times i.e. the red and green plots meet at late times. Black hole entropy is gravitational and  originates from the nonstationary interior region (see \cite{Edery_Constantineau} for a further discussion on this point). 

The function $F \eq -L$ at late times in fig. \ref{5DScalarGraAct} is equal to $F \eq 0.1764$. This needs to be compared to the free energy $E/3$ of a $5D$ Schwarzschild black hole of mass $E$, where $E$ is calculated via the relation $E=3\pi\,r_0^2/(2\,G_5)=9\,\pi^2\,r_0^2$ (this is then divided by $E_{tot}$ to obtain its value in ADM mass units). $E$ is plotted in fig. \ref{5DScalarGraAct} and at late times ($t=8$) its value is $0.5044$ so that $E/3 = 0.1682$. There is less than a $5\%$ difference between the numerical value of $F\eq-L$ at late times and the expected $E/3$ result.

%	\begin{figure}[tbp]
%		\begin{center}
%			\includegraphics[scale=.38, draft=false, trim=2cm 1.5cm 2.5cm 2cm, clip=true]{GraCumulTotal5D.eps}
%		\end{center}
%		\caption{\label{5DScalarGraCumulTotal} Accumulation plots for the total free energy for different times.  The % % % large dip at late times occurs immediately inside the event horizon where the metric field $\psi$ is nonstationary % % (see figure \ref{5DScalarGraPsiDot}).  The small dip between $10 \leq r \leq 20$ is due to the outgoing matter wave.}
%	\end{figure}
	
%	\begin{figure}[tbp]
%		\begin{center}
%			\includegraphics[scale=.38, draft=false, trim=2cm 1.5cm 2.5cm 2cm, clip=true]{GraChiDot5D.eps}
%		\end{center}
%		\caption{\label{5DScalarGraChiDot} Time dependence of the matter field $\chi$. $\chi$ is static except near the % % event horizon.}
%	\end{figure}

\section{Conclusion}

In this paper, we tracked numerically the negative of the Lagrangian, $-L$, during the spherical collapse of a massless scalar field to a black hole in isotropic coordinates. This was performed in 4D (3+1 dimensions) and in 5D and as far as we know, this is the first time that the physically relevant 3+1 case has been studied. We showed that at late stages of the collapse, $-L$ reached a numerical value close to the free energy $E/2$ for the 4D Schwarzschild black hole (to within $5\%$) and close to the free energy $E/3$ for the 5D case (again, to within 5\%). This confirms that previous 5D thermodynamic results from the collapse of a Yang-Mills instanton \cite{Khlebnikov} were universal and not dependent on the particular type of matter. We saw that the free energy stems almost entirely from the gravitational part $-L_G$ of the total Lagrangian: the matter contribution, $-L_{KG}$, was close to zero. The main reason for the $5\%$ discrepancy is that the fields have not evolved far enough in time before the total energy begins to deviate from unity. It should be possible to reduce the discrepancy in the future by increasing the resolution and implementing adaptive mesh refinement \cite{Oliger} to reduce simulation time.

Our 4D and 5D thermodynamic results obtained from the dynamics of classical gravitational collapse together with previous 5D results \cite{Khlebnikov} provide numerical evidence for associating a classical quantity, the negative of the gravitational Lagrangian $-L_G$, to the (Helmholtz) free energy $F\eq E-T\,S$ of a black hole. This opens up an interesting question. Can the product $TS$ have a classical interpretation without the temperature and entropy having a classical interpretation? It is sometimes stated that because black holes do not radiate classically, they must have zero temperature classically. However, it seems difficult to reconcile such a statement with what we know of the free energy i.e. that classically $TS \ne 0$. 

For a Schwarzschild black hole (vacuum solution), we showed analytically that the static exterior region makes a positive contribution $E$ to $-L_G$. It follows that the negative contribution $-T\,S$ to the free energy stems from the nonstationary interior region. This feature of the Schwarzschild black hole is observed in our numerical simulation. In our accumulation plot of $-L_G$, a negative contribution (a large dip) can be seen just inside the event horizon where the metric field $\psi$ is nonstationary. This ties in with Bekenstein's original view of black hole entropy as a measure of an outside observer's ignorance of the internal configurations hidden behind the event horizon \cite{Bekenstein}. As discussed in section 2.4.1, the internal configurations have recently been identified as points in phase space $[h_{ab},P^{ab}]$ \cite{Edery_Constantineau}. If the region behind the event horizon is nonstationary, then there is a continuous set of classical microstates $[h_{ab}(t),P^{ab}(t)]$ in the interior parametrized by the time $t$. An outside observer does not have access to the parameter $t$ and is clearly ignorant of which classical microstate the black hole interior is in.      

\begin{appendix}
\def\theequation{A.\arabic{equation}}
\setcounter{equation}{0}
\section{Derivation of the 5D equations of motion}

In this appendix we derive the 5D equations of motion. To remain general, we do not set here the mass $m$ of the scalar field to zero. To recover the equations of motion \reff{5DScalarEConstraint}-\reff{5DScalarEqN} simply set $m=0$. 

We begin with the Einstein field equations 
	\beq
		G_{\alpha\beta} = R_{\alpha\beta} - \frac{1}{2}R_5 \,g_{\alpha\beta} = 8\pi G_5T_{\alpha\beta} = \kappa^2T_{\alpha\beta}.
	\ee{}
The general expression for the stress-energy tensor $T_{\alpha\beta}$ of the KG field is  given by \reff{4DScalarStressEnergyTensor}. From the Lagrangian density of the 5D KG action, along with the metric tensor of the metric \reff{isometric5D}, we obtain
	\beq
		T_{\alpha\beta} = \partial_\alpha\chi \partial_\beta\chi - \frac{1}{2} \Big( \frac{\chi'^2}{\psi^2} - \frac{\dot{\chi}^2}{N^2} + m^2\chi^2 \Big)g_{\alpha\beta}.
	\ee{}
	By calculating the components of this tensor explicitly, one can observe that only seven of them, two off-diagonal and the five diagonal elements, are non-zero:
	\beqa
		T_{tt} &=& \frac{N^2}{2} \Big( \frac{\chi'^2}{\psi^2} + \frac{\dot{\chi}^2}{N^2} + m^2\chi^2 \Big) \,, \\
		T_{rr} &=& -\frac{\psi^2}{2} \Big( -\frac{\chi'^2}{\psi^2} - \frac{\dot{\chi}^2}{N^2} + m^2\chi^2 \Big) \,, \\
		T_{rt} &=& T_{tr} = \dot{\chi}\chi' \,, \\
		T_{\theta\theta} &=& -\frac{\psi^2r^2}{2} \Big( \frac{\chi'^2}{\psi^2} - \frac{\dot{\chi}^2}{N^2} + m^2\chi^2 \Big) \,, \\
		T_{\phi\phi} &=& \sin^2\theta \ T_{\theta\theta} \,, \\
		T_{\gamma\gamma} &=& \sin^2\theta\sin^2\phi \ T_{\theta\theta}.
	\eea{}
	
	The components of the Einstein tensor are
	\beqa
		G_{tt} &=& \frac{3}{r\psi^3} ( 2r\dot{\psi}^2\psi - 3\psi'N^2 - r\psi''N^2 ) \,, \\
		G_{rr} &=& \frac{-3}{r\psi^2N^3} ( -r\psi'^2N^3 - 2\psi'N^3\psi + r\dot{\psi}^2N\psi^2 - r\dot{N}\dot{\psi}\psi^3 \nonumber\\
		& & {} + r\ddot{\psi}N\psi^3 - rN'\psi'N^2\psi - N'N^2\psi^2 ) \,, \\
		G_{rt} &=& G_{tr} = \frac{-3}{\psi^2N} ( -\dot{\psi}\psi'N + \dot{\psi}'N\psi - \dot{\psi}N'\psi ) \,, \\
		G_{\theta\theta} &=& \frac{-r}{\psi^2N^3} ( 3r\dot{\psi}^2N\psi^2 - 4\psi'N^3\psi - 2r\psi''N^3\psi + r\psi'^2N^3 - 3r\dot{N}\dot{\psi}\psi^3 \nonumber\\
		& & {} + 3r\ddot{\psi}N\psi^3 - rN'\psi'N^2\psi - 2N'\psi^2N^2 - rN''N^2\psi^2 ) \,, \\
		G_{\phi\phi} &=& \sin^2\theta G_{\theta\theta} \,, \\
		G_{\gamma\gamma} &=& \sin^2\theta\sin^2\phi G_{\theta\theta}.
	\eea{}
	
The $\phi$, and $\gamma$ components of these two tensors can be expressed as a function of the $\theta$ component and the $rt$ and $tr$ components are equal, which leaves us with only four independent equations (the ones for the $tt$, $rr$, $rt$ and $\theta\theta$ components).
	
	\subsection{The energy constraint equation}
		The energy constraint equation is obtained from the equation $G_{tt} = \kappa^2 T_{tt}$, and is given here as
		\beq			
			\frac{3}{r\psi^3} ( 2r\dot{\psi}^2\psi - 3\psi'N^2 - r\psi''N^2 ) = \kappa^2 \frac{N^2}{2} \Big( \frac{\chi'^2}{\psi^2} + \frac{\dot{\chi}^2}{N^2} + m^2\chi^2 \Big).
		\eeq{5DScalarGtt2}
		This equation can be reduced to a more convenient form by using the flat space Laplacian \reff{5DScalarLaplacian} and the trace of the extrinsic curvature in 5D isotropic coordinates obtained from the equation
		\beqa
			\lefteqn{K = g^{ab}K_{ab} = \frac{-1}{2N} g^{ab}\partial_tg_{ab} = } \nonumber\\
			& & {} \frac{-1}{2N} \Big( \frac{2\psi\dot{\psi}}{\psi^2} + \frac{2\psi\dot{\psi}r^2}{\psi^2r^2} + \frac{2\psi\dot{\psi}r^2\sin^2\theta}{\psi^2r^2\sin^2\theta} + \frac{2\psi\dot{\psi}r^2\sin^2\theta\sin^2\phi}{\psi^2r^2\sin^2\theta\sin^2\phi} \Big),
		\eea{}
		which yields an evolution equation for the conformal factor $\psi$
		\beq			
			\frac{\dot{\psi}}{N} = -\frac{K\psi}{4}.
		\eeq{5DScalarEqPsi2}
		Dividing the energy constraint equation \reff{5DScalarGtt2} by $N^2$ and using the Laplacian \reff{5DScalarLaplacian} and \reff{5DScalarEqPsi2}, we obtain
		\beq
			\frac{3}{8}K^2 - 3\frac{\nabla^2\psi}{\psi^3} = \frac{\kappa^2}{2} \Big( \frac{\dot{\chi}^2}{N^2} + \frac{\chi'^2}{\psi^2} + m^2\chi^2 \Big).
		\ee{}
		The term in parenthesis is, as in 4D, interpreted as an energy density of the matter field $\chi$
		\beq
						\mathcal{E} = \frac{1}{2}\Big( \frac{\dot{\chi}^2}{N^2} + \frac{\chi'^2}{\psi^2} + V \Big),
		\eeq{5DScalarEDensityA2}
		where $V = m^2\,\chi^2$.
		
		The energy constraint equation in its final form is
		\beq			
			-3\frac{\nabla^2\psi}{\psi^3} = \kappa^2\mathcal{E} - \frac{3}{8}K^2.
		\eeq{5DScalarEConstraint2}
		
	\subsection{The momentum constraint equation}
		The momentum constraint equation is obtained from the Einstein equation $G_{rt} = \kappa^2T_{rt}$:
		\beq			
			\frac{-3}{\psi^2N} ( -\dot{\psi}\psi'N + \dot{\psi}'N\psi - \dot{\psi}N'\psi ) = \kappa^2 \dot{\chi}\chi' \,.
		\eeq{5DScalarEEqGrt2}
		From the spatial derivative of the evolution equation for $\psi$ \reff{5DScalarEqPsi2} we obtain
		\beq
			\frac{K'}{4} = -\Big( \frac{\dot{\psi}'}{N\psi} - \frac{N'\dot{\psi}}{N^2\psi} - \frac{\dot{\psi}\psi'}{N\psi^2} \Big).
		\ee{}
		By substituting the above equation in \reff{5DScalarEEqGrt2}, we obtain a more convenient form for the momentum constraint
		\beq
						\frac{K'}{4} = \frac{\kappa^2}{3}\frac{\dot{\chi}}{N}\chi'.
		\eeq{5DScalarMConstraintA2}
		
	\subsection{The evolution equation for $K$}
		The evolution equation for $K$ can be obtained by rearranging the Einstein equation $G_{rr} = \kappa^2T_{rr}$:
		\beqa			
			\frac{-3}{r\psi^2N^3} \Big( r\dot{\psi}^2N\psi^2 - 2\psi'N^3\psi 
			- r\psi'^2N^3 - r\dot{N}\dot{\psi}\psi^3 + r\ddot{\psi}N\psi^3 \nonumber\\- rN'\psi'N^2\psi - N'N^2\psi^2 \Big) =  \frac{\kappa^2\psi^2}{2} \Big( \frac{\chi'^2}{\psi^2} + \frac{\dot{\chi}^2}{N^2} - m^2\chi^2 \Big). \nonumber
		\eeqa{5DScalarEEqGrr2}
		The time derivatives terms in the left-hand side can be expressed in terms of $K$ and $\dot{K}$ from \reff{5DScalarEqPsi2}:
		\beq
					-3\frac{\dot{\psi}^2}{N^2} + 3\frac{\dot{N}\dot{\psi}\psi}{N^3} - 3\frac{\ddot{\psi}\psi}{N^2} = \frac{3\psi^2}{4} \Big( \frac{\dot{K}}{N} - \frac{K^2}{2} \Big).
		\eeq{5DScalarDotTermK2}
		Substituting the above equation and the definition of $\mathcal{E}$ \reff{5DScalarEDensityA2} in \reff{5DScalarEEqGrr2}, one obtains an evolution equation for the trace of the extrinsic curvature $K$:
		\beq		
			\frac{\dot{K}}{N} = \frac{K^2}{2} + \frac{4}{3}\kappa^2\mathcal{E} - \frac{4}{3}\kappa^2V - 4\frac{\psi'}{\psi^3} \Big( \frac{\psi'}{\psi} + \frac{2}{r} \Big) - 4\frac{N'}{N\psi^2} \Big( \frac{\psi'}{\psi} + \frac{1}{r} \Big).
		\eeq{5DScalarEqK2}
		
		From the prescription in \cite{Finelli}, we add a term proportional to the energy constraint \reff{5DScalarEConstraint2} to \reff{5DScalarEqK2} in order to get rid of numerical instabilities in the simulation:
		\beq
			-\frac{4}{3}wC_E = -\frac{4}{3}w \big( \kappa^2\mathcal{E} - \frac{3K^2}{8} + \frac{3\nabla^2\psi}{\psi^3} \big),
		\ee{}
		where $w$ is a positive constant that has to be unity or very close to it.  The final form of the evolution equation for $K$ is then
		\beqa			
			\frac{\dot{K}}{N} &=& \frac{K^2}{2} (1+w) + \frac{4}{3}\kappa^2\mathcal{E} (1-w) - \frac{4}{3}\kappa^2V \nonumber \\
			& & {} - 4\frac{\psi'}{\psi^3} \Big( \frac{\psi'}{\psi} + \frac{2}{r} \Big) - 4\frac{N'}{N\psi^2} \Big( \frac{\psi'}{\psi} + \frac{1}{r} \Big) - 4w\frac{\nabla^2\psi}{\psi^3}\,.
		\eeqa{5DScalarEEqK2}
		As $C_E = 0$ from \reff{5DScalarEConstraint2}, the additional term will not alter the mathematical content of the equation, so \reff{5DScalarEqK2} and \reff{5DScalarEEqK2} are equivalent in that respect.
		
	\subsection{An ODE for the lapse function}
		An ODE for the lapse function $N$ can be obtained from the last of Einstein's field equations, $G_{\theta\theta} = \kappa^2T_{\theta\theta}$, i.e.
		\beqa
					\frac{-r}{\psi^2N^3} ( r\psi'^2N^3 - 4\psi'N^3\psi - 2r\psi''N^3\psi & & \nonumber\\
			{} + 3r\dot{\psi}^2N\psi^2 - 3r\dot{N}\dot{\psi}\psi^3 - rN'\psi'N^2\psi \nonumber\\
			{} + 3r\ddot{\psi}N\psi^3 - 2N'^2\psi^2N^2 - rN''N^2\psi^2 ) &=& -\frac{\kappa^2\psi^2r^2}{2} \Big( \frac{\chi'^2}{\psi^2} - \frac{\dot{\chi}^2}{N^2} + m^2\chi^2 \Big).
		\eeqa{5DScalarGqq2}
		The combination of \reff{5DScalarDotTermK2} and \reff{5DScalarEqK2} yields
		\beqa
			\lefteqn{-3\frac{\dot{\psi}^2}{N^2} + 3\frac{\dot{N}\dot{\psi}\psi}{N^3} - 3\frac{\ddot{\psi}\psi}{N^2} =} \\
			& &  \psi^2 \big( \kappa^2\mathcal{E} - \kappa^2V \big) - 3 \Big( \frac{\psi'^2}{\psi^2} + \frac{2}{r}\frac{\psi'}{\psi} + \frac{N'}{N}\frac{\psi'}{\psi} + \frac{1}{r}\frac{N'}{N} \Big). \nonumber
		\eea{}
		Substituting the above equation into \reff{5DScalarGqq2}, one can obtain a convenient form for the ODE for $N$
		\beq		
			\frac{2r}{\psi}\partial_r \Big( \frac{\psi'}{r\psi^2} \Big) + \frac{r}{N}\partial_r \Big( \frac{N'}{r\psi^2} \Big) = -\kappa^2 \frac{\chi'^2}{\psi^2}.
		\eeq{5DScalarEqN2}

\subsection{Equations of motion for the matter field}
	The evolution equations related to the matter field $\chi$ and its ``conjugate momentum'' $p$ are now the only equations left to derive.  One way of obtaining them is from the Lagrange equation of motion
	\beq
		\frac{\partial L_{tot}}{\partial\chi} - \partial_\mu \Big( \frac{\partial L_{tot}}{\partial(\partial_\mu\chi)} \Big) = 0.
	\ee{}
	Substituting the total Lagrangian \reff{5DScalarLtot} and separating the spatial and time derivatives, this equation becomes
	\beq		
		4\frac{\dot{\psi}\dot{\chi}\psi^3}{N^2} + \frac{\ddot{\chi}\psi^4}{N^2} - \frac{\dot{N}\dot{\chi}\psi^4}{N^3} = \psi^2 \Big( \chi'' + \frac{N'}{N}\chi' + 2\frac{\psi'}{\psi}\chi' + \frac{3}{r}\chi' - \psi^2m^2\chi \Big).
	\eeq{5DScalarLEqChi2}
The left-hand side of this equation can be expressed as
	\beq
		4\frac{\dot{\psi}\dot{\chi}\psi^3}{N^2} + \frac{\ddot{\chi}\psi^4}{N^2} - \frac{\dot{N}\dot{\chi}\psi^4}{N^3} = \frac{1}{N}\partial_t \Big( \psi^4\frac{\dot{\chi}}{N} \Big).
	\ee{}
A new function $p$ is defined here as
	\beq
		p = \psi^4\frac{\dot{\chi}}{N}.
	\ee{}
This function can be roughly associated to the momentum conjugate to $\chi$. \reff{5DScalarLEqChi2} is then expressed by two single time derivative equations:
	\beq	
		\frac{\dot{\chi}}{N} = \frac{p}{\psi^4}
	\eeq{5DScalarEqChi2}
and
	\beq
			\frac{\dot{p}}{N} = \psi^2 \Big( \chi'' + \frac{N'}{N}\chi' + 2\frac{\psi'}{\psi}\chi' + \frac{3}{r}\chi' - \psi^2m^2\chi \Big).
	\eeq{5DScalarEqP2}
Furthermore, the evolution equation for $\chi$ \reff{5DScalarEqChi2} can be substituted in the energy density \reff{5DScalarEDensityA2} and in the momentum constraint \reff{5DScalarMConstraintA2} to obtain
	\beq
		\mathcal{E} = \frac{1}{2}\Big( \frac{p^2}{\psi^8} + \frac{\chi'^2}{\psi^2} + V \Big)
	\eeq{5DScalarEDensity2}
and
	\beq
		\frac{K'}{4} = \frac{\kappa^2}{3\,\psi^4}\,p\,\chi'.
	\eeq{5DScalarMConstraint2}
\end{appendix}

\section*{Acknowledgments}
A.E. acknowledges support from a discovery grant of the National
Science and Engineering Research Council of Canada (NSERC). B.C. 
acknowledges financial support from a Bishop's Senate Research Grant.

\end{document}